\documentclass[10pt]{article}
\usepackage{graphicx}
\usepackage{epsfig}
\usepackage{amsmath}

\usepackage{fancyheadings}
\newcommand{\elabel}[1]{\label{#1}}                         

\newcommand{\eg}{{e.g.\ }}

\newcommand{\ie}{{i.e.\ }}
\newcommand{\nonlocally}{non-locally}

\newcommand{\ams}{asymptotic multiple-scales}
\newcommand{\avg}[1]{\overline{{|#1|}^2}}
\newcommand{\cnls}{coupled NLS}
\newcommand{\cw}{coupled-wave}
\newcommand{\er}{e_r} 
\newcommand{\expf}[1]{{\rm e}^{#1}}

\newcommand{\nlcnls}{{\nonlocally} coupled NLS}
\newcommand{\nondis}{non-dispersive nonlinearly coupled envelope}
\newcommand{\sg}{Sine-Gordon}
\newcommand{\mydelta}{\sigma}
\newcommand{\superlong}{super-long}

\renewcommand{\Im}{{\rm Im}}

\newcommand{\eps}{\varepsilon}
\newcommand{\etabar}{{\overline \eta}}
\newcommand{\kcc}{\kd_{c_2}}           
\newcommand{\kd}{\hat k}               
\newcommand{\kdl}{\hat \kappa}         
\newcommand{\kdsl}{\hat K}             
\newcommand{\kn}{\hat k_n}   
\newcommand{\lx}{l}                    
\newcommand{\lxlong}{\lambda}          
\newcommand{\longx}{\chi}
\newcommand{\omegad}{\hat \omega}     
\newcommand{\omegadss}{\hat \Omega}   
\newcommand{\ps}{\hat v}              
\newcommand{\sd}{\hat s}              
\newcommand{\sn}{\hat s_n}                 
\newcommand{\xibar}{{\overline \xi}}
\newcommand{\qa}{Q_a}

\newcommand{\qc}{Q_c}

\newcommand{\amp}{U}
\newcommand{\amps}{\phi}
\newcommand{\ed}{\hat E}    
\newcommand{\eplane}{E_{plane}}      
\newcommand{\eside}[1]{E_{ side | {#1} }}       
\newcommand{\ec}{E_c}        
\newcommand{\eu}{E_u}         
\newcommand{\ew}{E_w}         
\newcommand{\vg}[1]{{v_g}_{#1}}
\newcommand{\vgbar}{\overline v_g}
\newcommand{\vgdev}{\Delta}

\begin{document}

\title{Modulational instability of two pairs of counter-propagating waves and energy exchange 
in two-component media}
\author {S.D. Griffiths$^a$, R.H.J. Grimshaw$^{b}$, K.R. Khusnutdinova$^{b}$
\footnote {Corresponding author.
  Tel.:\,+44 (0)1509 228202.  Fax:\,+44 (0)1509 223969.
   E-mail address: K.Khusnutdinova@lboro.ac.uk
   }}
\date{}
\maketitle

\thispagestyle{empty}
\vspace{-5ex}
\begin{center}
        $^a$Department of Atmospheric Sciences, University of Washington, \\
        Box 351640, Seattle, WA 98195, USA\\[3ex]
            $^b$Department of Mathematical Sciences,
        Loughborough University,\\
        Loughborough, LE11 3TU, UK  \\[5ex]
\end{center}

\begin{abstract}
The dynamics of two pairs of counter-propagating waves in two-component media
is considered within the framework of two generally nonintegrable coupled {\sg} equations.
We consider the dynamics of weakly nonlinear
wave packets, and using an {\ams} expansion we obtain a suite of
evolution equations to describe energy exchange
between the two components of the system.
Depending on the wave packet length-scale {\it vis-a-vis}
the wave amplitude scale, these evolution equations are either 
four non-dispersive and nonlinearly coupled envelope equations,
or four non-locally coupled nonlinear Schr\"odinger equations. 
We also consider a set of fully coupled nonlinear Schr\"odinger
equations, even though this system contains small dispersive terms which
are strictly beyond the leading order of the {\ams} expansion method.  
Using both the theoretical predictions following from these asymptotic models
and numerical simulations of the original unapproximated equations, 
we investigate the stability of plane-wave solutions, and show that they may be modulationally unstable. 
These instabilities can then lead to the formation of localized structures, and
to a modification of the energy exchange between the components.
When the system is close to being integrable, the time-evolution is distinguished by a remarkable almost periodic sequence of energy exchange scenarios, with spatial patterns alternating between approximately uniform wavetrains and localized structures. 
\end{abstract}

\vspace{5mm}



\vspace{5mm}


\newpage


\section{Introduction} \label{s:intro}

The dynamics of nonlinear waves is central to understanding the
behaviour of a wide range of physical systems. Applications include
Langmuir waves in plasmas (\eg~\cite{handbook}), waves in optical fibres (\eg~\cite{Agrawal,Kivshar}), and water waves (\eg~\cite{Yuen} and references therein).
An important theme of much of this work has been understanding modulational
instabilities of plane waves, and in recent years a particular focus has been
the instability of nonlinearly coupled waves. Examples
include studies of counter-propagating waves in parametrically
forced systems (\eg~\cite{Martel1,Martel2,Mancebo}), standing water waves
(\eg~\cite{Bridges}), co-propagating waves in shallow water~\cite{Onorato}, and more
general energy exchange processes in coupled {\sg} equations~\cite{Griffiths}. Other interesting recent results related to the dynamics of nonlinearly coupled waves include the rigorous proof of the existence of standing water waves~\cite{IPT}, and a study of resonant triad dynamics in weakly damped Faraday waves with two-frequency forcing~\cite{Porter}.  

Our present study is devoted to multi-component 
systems that may be modelled by nonlinear partial differential 
equations with one space dimension $x$, and with time $t$, which admit 
linear stable wave solutions of the form $A \exp{i(kx-\omega t)}$.
Here $\omega =\omega (k)$, where importantly this dispersion 
relation has several branches. 
We aim to study the nonlinear development of several plane
waves with the same wavenumber $k$, taking particular care to 
diagnose any modulational instabilities. As is customary we shall 
assume that each wave amplitude $A$ is  $O(\eps )$ where $\eps \ll 
1$, thus enabling the weakly nonlinear regime to be studied using
asymptotic methods. 

When there is just a single wave, the procedure is well-known. There is a
distinguished lengthscale for modulations to the wave amplitude, $O(\eps^{-1})$,
 on which the effects of dispersion and nonlinearity become simultaneously
important, on a timescale $O(\eps^{-2})$.  Then, in a reference frame moving 
with the group velocity of the wave,
the governing nonlinear partial differential equations can be reduced 
via the method of multiple scales to a nonlinear Schr\"odinger (NLS) 
equation for the wave amplitude $A$ (e.g. \cite{Newell}). 
As is well known, if $\omega''(k)$ has the same sign as
the coefficient of the cubic nonlinear term, then a plane wave is
modulationally unstable
\cite{Zakharov} (sometimes called a Benjamin--Feir instability, or sideband
instability \cite{Ben}).

One can apply a similar methodology for the case of two waves,
possibly counter-propagating, say $A \exp{i(kx-\omega_1(k) t)} + B
\exp{i(kx-\omega_2(k) t)}$. 
With the same scaling as above, in general the governing
 nonlinear partial differential equations can be 
reduced to {\nlcnls} equations for $A$ and $B$
\cite{Knobloch,Gibbon,Pierce}. For unbounded or periodic systems, the
coupling terms either disappear or can be eliminated by a phase 
transformation \cite{Gibbon}, and so modulational instability is 
determined by the instability of the individual waves. 
However, if the group velocities $\omega'_1$ and
$\omega'_2$ are equal, or differ only by an $O(\eps)$ quantity, 
then one obtains locally coupled NLS equations,
and there may be additional instabilities solely due to this coupling,
even when both waves are individually stable \cite{Forest1,Forest2}.

More generally, one can pursue an {\ams} expansion
in which the wave amplitudes, of $O(\eps)$, are modulated on a lengthscale
of $O(\mydelta^{-1})$. Then, retaining all terms of $O(\eps \mydelta ,\eps \mydelta^2
,\eps^3)$, the governing partial differential equations can be reduced
to {\cnls} equations for the two wave amplitudes. Expressed here
in unscaled variables, they are of the form
 \begin{equation}
  A_t + \vg{1} A_x = (i \omega''_1/2) A_{xx} 
      + i \mu_1 |A|^2 A + i \mu_2 |B|^2 A,
  \elabel{e:cnls}
 \end{equation}
where $\vg{1}=\omega'_1$ is the group velocity,
and $\mu_{1,2}$ are constant nonlinear coefficients. There is a corresponding equation for
$B$, with a different transport term due to the group
velocity $\vg{2}=\omega'_2$. On the right-hand side, the linear
dispersive terms of $O(\eps \mydelta^2 )$ are retained, even though
these are strictly smaller than the 
$O(\eps \mydelta )$ transport terms on the left-hand side. 
Note that for a single wave, the transport term can be removed
by moving to the group velocity reference frame, but this is generally
not possible in coupled systems with differing group velocities.
The nonlinear terms on the right-hand side are $O(\eps^3)$, 
and with the conventional NLS choice that $\mydelta =\eps$,
these equations can then be systematically reduced to the 
aforementioned non-locally coupled NLS equations,
provided that $\vg{1}$ is not equal or close to $\vg{2}$.
Alternatively, one could choose $\mydelta = \eps^2 $ giving a balance
between the transport terms and the nonlinear terms, leading to
non-dispersive  and nonlinearly coupled envelope equations.
Nevertheless, although {\cnls} equations of the form (\ref{e:cnls}) are not 
generally asymptotically exact, in that they contain small dispersive terms
which are strictly beyond the leading order transport terms,
they may be useful as they do capture the leading order dispersive
and nonlinear effects.
Studies based on this type of approach have recently been made for 
the parametric
generation of waves by an external oscillatory field in the framework
of two coupled, damped, parametrically driven NLS equations (see
\cite{Martel1,Martel2,Mancebo} and references therein), and for the
interaction of two waves in shallow water in the
framework of the KdV equation \cite{Onorato}. 

The aims of this study are to extend our understanding
of the nonlinear behaviour of multiple-wave systems, to derive
asymptotic equations to describe many interacting waves, and to assess
the applicability and accuracy of the various resulting coupled wave
models, with particular reference to the modulational
instability of plane waves. 
This work is performed in the context of energy exchange processes in
two-component systems, permitting us to examine explicitly both 
two-wave and four-wave dynamics. We consider a system of coupled 
nonlinear Klein--Gordon equations
   \begin{equation}
    u_{tt}  -  u_{xx}  =  f_u (u,w), \quad
    w_{tt}  -  c^2 w_{xx}  =  f_w (u,w), \elabel{1.1}
\end{equation}
where the subscripts denote partial derivatives. Here, $f(u,w)$ is a
potential function for the nonlinear coupling, and $c$ is the ratio of the
acoustic velocities of the components $u$ and $w$.
In its simplest form, (\ref{1.1}) describes the long-wave dynamics of
two coupled one-dimensional periodic chains of particles, the
elements of each chain being linked by  linear coupling, and the
chains themselves being linked by the nonlinear coupling \cite{Akh}.
Applications of this model include nonlinear waves in bi-layers with
imperfect interfaces \cite{Khus1,Khus2}, and dynamical processes in
DNA \cite{Yom,Yak}.

The particular case we shall study is that for which the 
potential function is
$f(u, w) = \cos (\delta u - w) - 1$. Then the system (\ref{1.1}) 
reduces to a set of coupled {\sg} equations:
   \begin{equation}
    u_{tt}  -  u_{xx}  =  - \delta^2 \sin (u-w), \quad
    w_{tt}  -  c^2 w_{xx}  =  \sin (u-w), \elabel{1.2}
   \end{equation}
where the variable $u$ replaces $\delta u$, compared to
(\ref{1.1}). In addition to the two applications mentioned above, the
coupled {\sg} equations generalize the Frenkel--Kontorova dislocation
model \cite{FK1} (see also \cite{Bra}, and references therein).
As shown in \cite{Khus}, processes involving an exchange of energy
between the two components of the system are possible due to the
existence of two branches of the dispersion relation.
These processes represent a continuum generalization
of energy exchange in a system of coupled pendulums \cite{Man} (see
also \cite{Arn}).  The mechanism of such energy exchange is essentially linear,
with energy being transferred between the physical components of the
system, rather than the wave components.
Therefore, a natural question to ask is, what happens to these linear 
plane wave solutions under the influence
of nonlinearity,  especially if they are modulationally perturbed?
Here, we are interested in studying such modulational instabilities in a more or less generic situation, in contrast to the recent papers on targeted energy transfer, which placed certain restrictions on the physical properties of the system \cite{TET1,TET2}.

First steps in this direction have been made in our recent work
\cite{Griffiths}. There we dealt with a particular solution of~(\ref{1.2}) involving
only  a single pair of waves, and considered its modulational instability in the context of  two  {\nlcnls} equations. 
In the present study, we generalise this analysis to
two pairs of counter-propagating waves, which generally leads to a
system of four {\nlcnls} equations. Note that the four-wave case 
is essentially different from the two-wave case, since the number of 
characteristic variables now differs from the number of independent 
variables, making the strict derivation of the asymptotic evolution 
equations not so straightforward.
However, as we show here,  the same methodology can, in
general, be used to derive $n$ {\nonlocally} coupled NLS equations for 
the co-evolution of $n$ waves.  We also re-examine
the two-wave stability problem using the {\cnls} equations,
and compare the predictions with those of the {\nlcnls} equations. 
Throughout, numerical
simulations of the governing coupled {\sg} equations  
are used to assess the various predictions over a range of 
wave amplitudes.

Our paper is organized as follows. In Section~\ref{s:exchange}, we
discuss solutions exhibiting periodic energy exchange
between the two components. Section~\ref{s:models} is devoted to the
derivation of asymptotic models for weakly nonlinear waves. 
Specifically, we derive {\nlcnls} equations for the 
regime with modulations on a lengthscale of $O(\eps^{-1})$, and also
{\nondis} equations for the regime with modulations on a lengthscale of 
$O(\eps^{-2})$. Then we motivate the introduction of the {\cnls} 
equations, and give them for the case of two pairs of counter-propagating waves (correcting an error in \cite{Khus}).
In Section~\ref{s:2wave}, we study modulational 
instability of weakly nonlinear coupled two-wave solutions in each of the 
respective models, extending our previous study \cite{Griffiths}. 
In Section~\ref{s:4wave}, we study modulational instability of weakly nonlinear four-wave solutions, using two asymptotic models.
Numerical simulations of the system (\ref{1.2}) of coupled {\sg}
equations are presented in Section~\ref{s:numerics}, with a focus on the
growth of any modulational instabilities, and the numerical results
are compared with the theoretical predictions of
Sections~\ref{s:2wave} and~\ref{s:4wave}. We conclude in Section~\ref{s:conclu}.


\section{Energy exchange} \label{s:exchange}
 
The system (\ref{1.1}) is Lagrangian, with density
 \begin{equation*}
   L = \frac12 (u_t^2 + w_t^2 - u_x^2 - c^2 w_x^2) + f(u, w).
 \end{equation*}
For any function $f(u, w)$, the system (\ref{1.1}) has conservation 
laws for energy and momentum.
(In particular cases, there are additional conservation laws, which 
can be found using results from a Lie group classification of these 
equations~\cite{Khus1},  and by using the Noether theorem (e.g.\ \cite{Olver})).

The energy conservation law for the system (\ref{1.2}) has the form
  \begin{equation*}
   \frac{\partial}{\partial t} \left( \frac{1}{2}
     \left( u_t^2 + u_x^2 + \delta^2 w_t^2 + c^2 \delta^2 w_x^2 \right)
      + \delta^2 \left( 1 - \cos(u-w) \right) \right)
   - \frac{\partial}{\partial x} \left(
        u_t u_x + c^2 \delta^2 w_t w_x \right) = 0.
  \end{equation*}
For periodic solutions in $0<x<\lx$, we define domain integrated 
average energies
  \begin{equation}
    \eu = \frac{1}{\lx} \int_0^{\lx}
      \frac{1}{2} \left( u_t^2 + u_x^2 \right) {\rm d} x, \quad
    \ew = \frac{1}{\lx} \int_0^{\lx}
      \frac{\delta^2}{2} \left( w_t^2 + c^2 w_x^2 \right) {\rm d} x,
   \elabel{euw}
  \end{equation}
  \begin{equation}
    \ec = \frac{1}{\lx} \int_0^{\lx}
          \delta^2 \left( 1 - \cos(u-w) \right) {\rm d} x,
     \elabel{ec}
  \end{equation}
describing the energy in the $u$-component, the energy in the 
$w$-component, and the energy of coupling, respectively. Then the 
domain integrated average energy $E$ is simply
  \begin{equation}
      E = \eu + \ew + \ec.
   \elabel{e}
  \end{equation}
  Although $E$ is constant, we can identify solutions
  that have a time-periodic exchange of energy between $\eu$, $\ew$ and
  $\ec$.
\subsection{Linear solutions} \label{s:linear}
When $|u-w| \ll 1$, the coupled {\sg} equations (\ref{1.2}) reduce to a
linear system. Looking for a solution of the form
$(u,w)=(1,\alpha) \exp{i(kx-\omega t)} + {\rm c.c.}$
leads to the dispersion relation
     \begin{equation}
      \omega_{1,2}^2 = \frac{1}{2} \left[ \nu_1^2 + \nu_2^2 \mp \sqrt{(\nu_1^2
      - \nu_2^2)^2 + 4 \delta^2} \right],
      \elabel{omega12}
     \end{equation}
where $\nu_1^2 = \delta^2 + k^2$ and $\nu_2^2 = 1 + c^2 k^2$.
The corresponding values of $\alpha$, giving the ratio $w/u$, are
\begin{equation*}
      \alpha_{1,2} = \frac{\nu_1^2 - \omega_{1,2}^2}{\delta^2}
                   = \frac{1}{\nu_2^2 - \omega_{1,2}^2}.
\end{equation*}
We note that $\alpha_1 >0$, and $\alpha_2 < 0$.
For example, if $c=1$, then $\alpha_1 = 1$ and $\alpha_2 = - \delta^{-2}$.

We first consider a superposition of two linear waves, taken to be
right-propagating, with the same positive wavenumber $k$ but
different frequencies $\omega_1(k)$ and $\omega_2(k)$. Further, we
suppose that one component, taken to be $u$, is initially excited,
while the other component $w$ is initially at rest. The corresponding  initial conditions are
 \begin{equation}
   \begin{split} 
    u(x,0) = \amp \cos kx, \; \; \; \;
    & u_t(x,0) = \frac{ 
                     \left( \omega_1 |\alpha_2| + \omega_2 \alpha_1 \right) }
                           {\alpha_1+|\alpha_2|} \, \amp \sin kx, \\
     w(x,0) = 0, \; \; \; \;
     & w_t(x,0)= \frac{ \left( \omega_1-\omega_2 \right) \alpha_1 |\alpha_2|  }
                           {\alpha_1+|\alpha_2|}  \, \amp \sin kx, 
   \end{split}
  \elabel{ics2} 
 \end{equation} 
and one may write the time-dependent  solution as
   \begin{equation}
      u = \tilde U(t) \, \cos (kx - \gamma_+ t - \theta(t) ), \quad
      w = - \frac{2 \alpha_1 |\alpha_2|}{\alpha_1 + |\alpha_2|}
                \, \amp \sin \gamma_{-} t\ \sin (kx - \gamma_{+} t),
     \elabel{2.3}
   \end{equation}
where
   \begin{equation}
    \tilde U(t) = \amp \sqrt{1 - \frac{4 \alpha_1 |\alpha_2| \sin^2 \gamma_- t}
                              {(\alpha_1+|\alpha_2|)^2}}, \quad
    \theta(t) = \arctan \left( \left( \frac{\alpha_1 - |\alpha_2|}
                {\alpha_1 + |\alpha_2|} \right) \tan \gamma_- t \right), \quad
    \gamma_{\pm} = \frac{\omega_2 \pm \omega_1}{2}. 
    \elabel{2.3b}
   \end{equation}
As discussed in more detail in~\cite{Griffiths}, this solution
describes an exchange of energy between waves travelling in the two
components $u$ and $w$. The amplitude of the wave propagating in the
$u$ component varies between its initial value $\amp$, and a
smaller non-zero value 
$\amp_{min}=\amp \left| \alpha_1 - \left| \alpha_2 \right| \right| / 
(\alpha_1 + |\alpha_2|)$.
There is a full exchange in the wave amplitudes from zero to non-zero
   values when
   \begin{equation}
     (c^2-1)k^2 = \delta^2 - 1.
    \elabel{2.3a}
   \end{equation}
Then $\nu_1^2 = \nu_2^2 = \nu^2$, $\alpha_1 = -\alpha_2 =
\delta^{-1}$, and the solution (\ref{2.3}) reduces to
   \begin{equation}
     u = \amp \, \cos \ \gamma_{-} t\ \cos (kx - \gamma_{+} t), \; \; \;
     w = -\amp \, \delta^{-1} \sin \ \gamma_{-} t\ \sin (kx - \gamma_{+} t).
    \elabel{2.4}
   \end{equation}
This is illustrated in Figure 1, for the case $c=\delta=1, k = 1.6$.
If $c=1$, then (\ref{2.3a}) can be satisfied only for $\delta=1$, while $k$
is arbitrary. If $c \neq 1$, then (\ref{2.3a}) cannot be satisfied if $(\delta^2-1)/(c^2-1) < 0$. Otherwise, (\ref{2.3a}) is satisfied for a single wavenumber $k = ((\delta^2-1)/(c^2-1))^{1/2}$.

\begin{figure}[htbp]
\begin{center}
\begin{picture}(160,40)(0,0)
\put(0,0){\includegraphics[width=140mm]{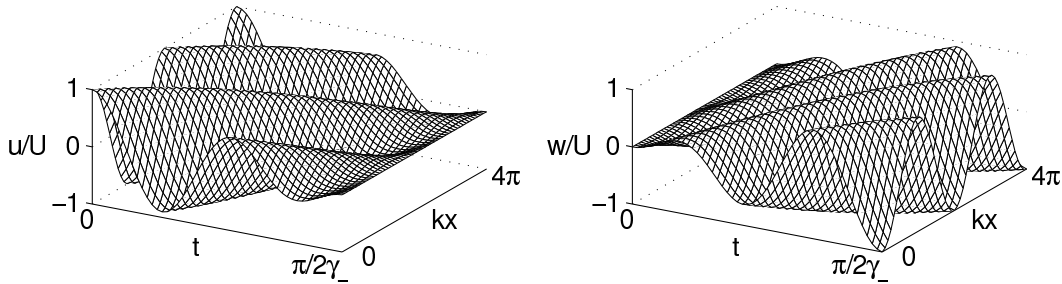}}
\end{picture}
\end{center}
\vspace*{-5mm}
\caption{{\small Full exchange in the linear two-wave solution (\ref{2.4}).}}
\label{fig1}
\end{figure}

For the solution (\ref{2.4}), the energy is partitioned according to
  \begin{equation}
   \begin{pmatrix} \eu \\ \ew \\ \ec \end{pmatrix} =
    \frac{\amp^2 \cos^2 \gamma_- t}{4}  
    \begin{pmatrix}
       \gamma_+^2 + k^2 \\ \gamma_-^2 \\ \delta^2 
    \end{pmatrix}  
  + \frac{\amp^2 \sin^2 \gamma_- t}{4} 
     \begin{pmatrix}
       \gamma_-^2 \\  \gamma_+^2 + c^2 k^2 \\ 1 
     \end{pmatrix},
   \elabel{2.5}
  \end{equation}
where, for these linear solutions, $\ec$ is calculated using only the
quadratic term of (\ref{ec}). Using $\gamma_-^2+\gamma_+^2 =
\delta^2+k^2$, one may show from (\ref{e}) and (\ref{2.5}) that 
$E=U^2 \left( \delta^2 + k^2
\right) / 2$. From (\ref{2.5}) we see that there is a partial 
exchange of energy between $u$ and
  $w$, with period $T=\pi/\gamma_-$. However, if $\gamma_-$ is small
  enough, \ie if the frequencies $\omega_1$ and $\omega_2$ are close
  enough for some wavenumber $k$, then at $t=\pi/2 \gamma_-, 3 \pi/2 
\gamma_-, \cdots$ the $u$-component is almost in equilibrium and 
almost all of its energy is lost, whilst at $t=0, \pi/ \gamma_{-}, 
\cdots$ the $w$-component is almost in equilibrium and almost all of 
its energy is lost. If $\nu^2 \gg \delta$, then the period $T$ of the 
energy exchange tends to infinity.

\begin{figure}[htbp]
\begin{center}
\begin{picture}(160,40)(0,0)
\put(0,0){\includegraphics[width=140mm]{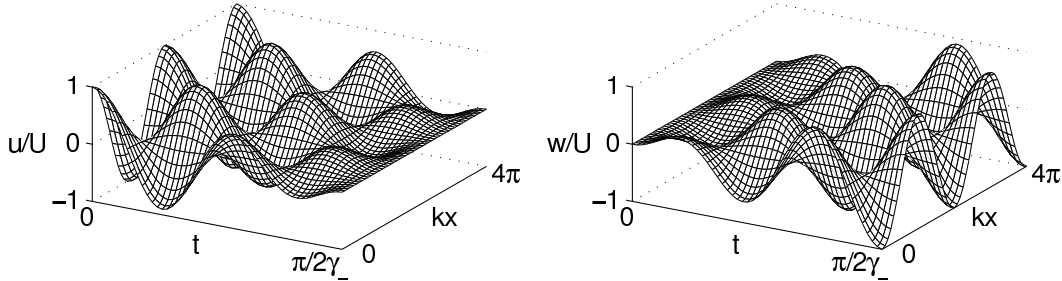}}
\end{picture}
\end{center}
\vspace*{-5mm}
\caption{{\small Full exchange in the linear four-wave solution (\ref{2.7}).}}
\label{fig2}
\end{figure}

There exists a corresponding solution to (\ref{2.3}) consisting of a
superposition of two left-propagating waves. A combination of this left-propagating solution and the right-propagating solution (\ref{2.3}) is obtained by using the initial conditions
\begin{equation}
   u(x,0) = \amp \cos kx, \; \;
   u_t(x,0) = 0, \; \;
   w(x,0) = 0, \; \; w_t(x,0)=0.
   \elabel{ics4} 
 \end{equation} 
We then obtain a standing wave involving two pairs of counter-propagating waves:
   \begin{equation}
    u = \tilde U(t) \cos (\gamma_+ t + \theta(t) ) \cos kx, \quad
    w = \frac{2 \alpha_1 |\alpha_2| \amp}{\alpha_1 + |\alpha_2|}
           \sin \ \gamma_{-} t\ \sin \ \gamma_{+} t\ \cos kx,
           \elabel{2.6}
   \end{equation}
with $\tilde U(t)$, $\theta(t)$ and $\gamma_{\pm}$ as  given by (\ref{2.3b}). 
The amplitude of the standing wave in the
$u$ component varies between $\amp$ and $\amp_{min}$, so that 
once again there is only a partial exchange between the amplitude of the components, unless condition (\ref{2.3a}) is satisfied. Then, we have a full exchange, and (\ref{2.6}) becomes
   \begin{equation}
    u = \amp \cos \ \gamma_{-} t\ \cos \ \gamma_{+} t\ \cos kx, \quad
    w = \amp \, \delta^{-1} \sin \ \gamma_{-} t\ \sin \ \gamma_{+} t\ \cos kx,
    \elabel{2.7}
   \end{equation}
where we note that $\gamma_{-} < \gamma_{+}$, from (\ref{2.3b}).
This standing wave solution is illustrated in Figure~\ref{fig2}, for
    the case $c=\delta=1$, $k=1.6$.
    
The energy in solution (\ref{2.7}) is $E=\amp^2 (\delta^2 + k^2)/4$, 
although its partition is rather more complicated than (\ref{2.5}). 
Nevertheless, the general pattern of a partial energy 
exchange on a timescale $\pi/\gamma_-$  remains.
For instance, for $\gamma_- \ll 1$ we can say that $u$ loses 
almost all of its energy near $t=\pi/2\gamma_-, 3\pi / 2 \gamma_-, 
\cdots$ etc.  Some examples of this exchange will be given in the 
numerical simulations of Section~\ref{s:numerics}.

Note that the role of $t$ and $x$ is interchangeable here; one can consider
a superposition of waves with one and the same frequency, but
different wavenumbers.

\subsection{Nonlinear solutions}
An exact nonlinear two-wave solution describing energy
exchange between the two components $u$ and $w$ of the system
(\ref{1.2}) can be found if $c = 1$, i.e., if the characteristic
velocities of the components coincide \cite{Khus}. It is given in terms of
Jacobi elliptic functions:
   \begin{equation}
    \begin{pmatrix} u \\ w \end{pmatrix} =
       \frac{2}{1+\delta^2} \left(
          {\rm arcsin}\ \phi_1 \begin{pmatrix} 1 \\ 1 \end{pmatrix}
        + {\rm arcsin}\ \phi_2 \begin{pmatrix} \delta^2 \\ -1 \end{pmatrix}
       \right), \quad
     \phi_{1,2} = \kappa \ {\rm sn} (k x - \omega_{1,2}(k) t + \theta_0, \kappa),
    \elabel{3.1}
   \end{equation}
where $\omega_1=k$, $\omega_2=\sqrt{1+\delta^2+k^2}$, from (\ref{omega12}). 
When $\delta^2 = 1$ (i.e.\ $\nu_1^2 = \nu_2^2$), the nonlinear solution
(\ref{3.1}) describes a full periodic exchange between the
two components of the system (see \cite{Khus} for details).

The form of the solution for small amplitude can be calculated by
taking the limit $\kappa \rightarrow 0$ in (\ref{3.1}). Using the
well-known approximation (e.g.\ \cite{Abram})
   \begin{displaymath}
     {\rm sn}(x, \kappa) = \sin x
              - \frac 14 \kappa^2 (x - \sin x \cos x) \cos x + O(\kappa^4),
   \end{displaymath}
one can eliminate the secular terms by renormalizing the frequencies
$\omega_{1,2}(k)$ in the solution (\ref{3.1}), and thus find an asymptotic
expansion with leading order terms:
  \begin{multline}
    \begin{pmatrix} u \\ w \end{pmatrix} =
       \frac{2 \kappa}{1+\delta^2} \left(
          \sin \left( kx - \omega_1(k) t + \tilde \theta_0 \right)
          \begin{pmatrix} 1 \\ 1 \end{pmatrix} \right. \\
         \left. 
        + \sin \left( kx - \omega_2(k) t + \tilde \theta_0
            + \Omega_2 \kappa^2 t + O(\kappa^4) \right)
          \begin{pmatrix} \delta^2 \\ -1 \end{pmatrix}
       \right) + O(\kappa^3),
    \elabel{3.2}
   \end{multline}
where $\Omega_2 = (1+\delta^2)/ 4 \omega_2$. Setting $\kappa=U/2$ and
$\tilde \theta_0=\pi/2$, the leading order terms of (\ref{3.2}) reduce to
the linear solution (\ref{2.3}) with $c=1$. 
 
In the general case, when $c \ne 1$, and for a greater number of
waves, such exact solutions are not generally available.
Instead, we study weakly nonlinear solutions by deriving appropriate
asymptotic models. We do this in the next section, and return to these
particular energy exchange solutions later.


\section{Models for weakly nonlinear wave packets} \label{s:models}

We consider the weakly nonlinear evolution of pairs of
counter-propagating plane waves within system (\ref{1.2}).
To do this we derive equations for the evolution of wave packets, each
with the same dominant wavenumber $k$. We suppose that the wave
amplitudes are characterised by a small parameter $\eps \ll 1$, and
thus introduce the {\ams} expansion
   \begin{displaymath}
    \begin{pmatrix} u \\ w \end{pmatrix} =
         \eps \begin{pmatrix} u_1 \\ w_1 \end{pmatrix}
       + \eps^2 \begin{pmatrix} u_2 \\ w_2 \end{pmatrix}
       + \eps^3 \begin{pmatrix} u_3 \\ w_3 \end{pmatrix} + O(\eps^4),
   \end{displaymath}
where
   \begin{equation}
    \begin{pmatrix} u_1 \\ w_1 \end{pmatrix} = \left[ A \expf{i(k x - \omega_1 t)}
                     + B \expf{i(k x + \omega_1 t)}\right]
                       \begin{pmatrix} 1 \\ \alpha_1 \end{pmatrix}
                      + \left[ C \expf{i(k x - \omega_2 t)}
                      + D \expf{i(k x + \omega_2 t)}\right]
                       \begin{pmatrix} 1 \\ \alpha_2 \end{pmatrix}
                      + {\rm c.c.} \elabel{uw1}
   \end{equation}
The coupled {\sg} equations (\ref{1.2}) may then be reduced to coupled
equations for the slow spatio-temporal evolution of the wave
amplitudes $A$, $B$, $C$ and $D$, which are functions of slow
variables. We will consider modulations on a long lengthscale $\longx$
or on a {\superlong} lengthscale $X$, with corresponding slow and
super-slow timescales $\tau$ and $T$, defined by
  \begin{equation}
   \longx = \eps x, \quad
   X = \eps^2 x, \quad
   \tau = \eps t, \quad
   T = \eps^2 t.
  \elabel{scales}
  \end{equation}
We first derive an asymptotic model for modulations on the 
{\superlong} lengthscale, leading to a system of four {\nondis} 
equations. We then do the same for modulations on the long
lengthscale, generally leading to a system of four {\nlcnls} equations
(``averaged equations''). We finally present a third model, the
{\cnls} equations, which may be regarded as a model for modulations on
the long lengthscale with both the $O(\eps^2)$ transport terms and the 
$O(\eps^3)$ linear dispersive terms retained.
 
\subsection{The {\nondis} equations} \label{s:superlong}

To derive this model,  one has to suppose that $A$, $B$, $C$ and $D$ are 
functions of $X = \eps^2 x, T=\eps^2 t$ (see, for example, \cite{Benney,AS}).
At leading order, \ie $O(\eps)$ in this expansion, we
find a linear wave equation for $u_1$ and $w_1$, which is taken to be
the four-wave solution (\ref{uw1}), with $A$, $B$, $C$ and $D$ as yet
undetermined functions of $X,T$.
At the next order, one obtains
solvability conditions in the usual way from the equations for $u_2$
and $w_2$, yielding
  \begin{equation}
    \begin{split}
    & A_T + \vg{1} A_X = i \mu_1 (|A|^2 + 2 |B|^2) A + i \mu_2 (|C|^2 
+ |D|^2)A + i \mu_2 B^* C D, \\
    & B_T - \vg{1} B_X =
     - i \mu_1 (|B|^2 + 2 |A|^2) B - i \mu_2 (|C|^2 + |D|^2)B - i \mu_2 A^* CD, \\
    & C_T + \vg{2} C_X =  i \mu_3 (|C|^2 + 2 |D|^2) C + i \mu_4 (|A|^2 
+ |B|^2)C + i \mu_4 D^* AB, \\
    & D_T - \vg{2} D_X =
     - i \mu_3 (|D|^2 + 2 |C|^2) D - i \mu_4 (|A|^2 + |B|^2)D - i \mu_4 C^* AB,
    \end{split}
   \elabel {ABCDeq1}
  \end{equation}
where the group velocities $\vg{i}$ are given by
  \begin{equation}
     \vg{i} = \omega'_i = \frac {k}{\omega_i}
               \frac{1 + \alpha_i^2 \delta^2 c^2}{1 + \alpha_i^2
\delta^2},
    \elabel{vg}
   \end{equation}
and the nonlinear coefficients are
  \begin{equation}
   \begin{split}
    & \mu_1 = \frac{\delta^2 (1 - \alpha_1)^4}
                {4 \omega_1 (1 + \alpha_1^2 \delta^2)}, \
       \mu_2 = \frac{\delta^2 (1 - \alpha_1)^2 (1 - \alpha_2)^2}{2 \omega_1 (1 +
       \alpha_1^2 \delta^2)},  \\
    & \mu_3 = \frac{\delta^2 (1 -
        \alpha_2)^4}{4 \omega_2 (1 + \alpha_2^2 \delta^2)}, \
        \mu_4 = \frac{\delta^2 (1 - \alpha_1)^2 (1 - \alpha_2)^2}{2 
\omega_2 (1 +
        \alpha_2^2 \delta^2)}.
    \elabel{mu}
  \end{split}
\end{equation}
The equations (\ref{ABCDeq1}) are non-dispersive. 

Note that for the case of  two-wave interaction, with $B=D=0$ for instance,  (\ref{ABCDeq1}) yields the system
\begin{equation*}
A_T + \vg{1} A_X = i (\mu_1 |A|^2  +  \mu_2  |C|^2) A, \hspace*{10mm} 
C_T + \vg{2} C_X = i (\mu_3 |C|^2 +  \mu_4 |A|^2) C,
\end{equation*}
which is integrable in quadratures \cite{Griffiths} (see also \cite{Benney}). 

\subsection{The {\nlcnls} equations} \label{s:long}

To derive the second model, we generalise the approach of
\cite{Knobloch} to the case of four waves. The generalization is 
not straightforward, since the number of characteristic variables now 
differs from the number of independent variables, whereas they are the same 
in the two-wave case. However, these potential
difficulties can be readily overcome, and
 the approach we describe here can generally be 
applied to any number of waves.

We suppose that $A$, $B$, $C$ and $D$ are functions of the long
lengthscale $\longx=\eps x$, in which case they must also depend on
both slow timescales $\tau = \eps t$ and $T = \eps^2 t$.
Once again, at $O(\eps)$ (\ie the leading order in this expansion), 
we find a linear wave equation for $u_1$ and $w_1$, which
is taken to be the four-wave solution (\ref{uw1}), with $A$, $B$, $C$
and $D$ as yet undetermined functions of $\longx$, $\tau$ and $T$. At
$O(\eps^2)$, there are solvability conditions $A_{\tau} + \vg{1}
A_{\longx} = 0$, $B_{\tau} - \vg{1} B_{\longx} = 0$, $C_{\tau} +
\vg{2} C_{\longx} = 0$, and $D_{\tau} - \vg{2} D_{\longx} = 0$.
Introducing the variables
   \begin{equation}
      \eta_i = \longx - \vg{i} \tau, \quad
        \xi_i = \longx + \vg{i} \tau,
     \elabel{etaxi}
   \end{equation}
the solvability conditions give
   \begin{equation}
     A = A(\eta_1, T), \quad B = B(\xi_1, T), \quad C = C(\eta_2, T),
     \quad D = D(\xi_2, T),
    \elabel{ABCDform1}
   \end{equation}
\ie there is a simple functional dependence for each wave amplitude
in its own
reference frame. If we suppose that our solutions have spatial period
$\lx$ in $x$, which must satisfy $\lx \sim \eps^{-1}$ for 
modulations on the long lengthscale, then $A, B, C$ and $D$ are periodic in 
$\eta_1, \xi_1, \eta_2$ and $\xi_2$ respectively, with period 
$\lxlong=\eps \lx$.

The $O(\eps^2)$ terms in the expansion may then be written as
   \begin{multline}
     \begin{pmatrix} u_2 \\ w_2 \end{pmatrix} = \left [
          A_2(\eta_1,\tau,T) \expf{i(kx - \omega_1 t)}
        + B_2(\xi_1, \tau, T)\expf{i(kx + \omega_1 t)}
                           \right ]  \begin{pmatrix} 1 \\ \alpha_1 
\end{pmatrix}\\
                       + \left [
          C_2(\eta_2,\tau,T) \expf{i(kx - \omega_2 t)}
        + D_2(\xi_2,\tau, T) \expf{i(kx + \omega_2 t)}
                           \right ] \begin{pmatrix} 1 \\ \alpha_2 
\end{pmatrix} + {\rm c.c.}
    \elabel{uw2}
   \end{multline}
At $O(\eps^3)$, one obtains solvability conditions in the usual way:
    \begin{align}
     \partial_\tau A_2 + A_T
              & = \frac{i}{2} \omega_1'' A_{\eta_1 \eta_1}
                  + i \mu_1 (|A|^2 + 2 |B|^2) A + i \mu_2 (|C|^2 + |D|^2) A + i\mu_2 B^* CD, 
                  \elabel{ABCDeq2a} \\
     \partial_\tau B_2 + B_T
              & = -\frac{i}{2} \omega_1'' B_{\xi_1 \xi_1}
                  - i \mu_1 (|B|^2 + 2 |A|^2) B - i \mu_2 (|C|^2 + |D|^2) B - i\mu_2 A^* CD, 
                  \elabel{ABCDeq2b} \\
     \partial_\tau C_2 + C_T
              & = \frac{i}{2} \omega_2'' C_{\eta_2 \eta_2}
                  + i \mu_3 (|C|^2 + 2 |D|^2) C + i \mu_4 (|A|^2 + |B|^2) C + i \mu_4 D^* AB, 
                  \elabel{ABCDeq2c} \\
     \partial_\tau D_2 + D_T
              & = -\frac{i}{2} \omega_2'' D_{\xi_2 \xi_2}
                  - i \mu_3 (|D|^2 + 2 |C|^2) D - i \mu_4 (|A|^2 + |B|^2) D - i\mu_4 C^* AB.
     \elabel{ABCDeq2d}
   \end{align}
Here the partial derivatives of $A_2$, $B_2$, $C_2$ and $D_2$ with
respect to $\tau$ are evaluated at constant $\eta_1$, $\xi_1$,
$\eta_2$ and $\xi_2$ respectively, consistent with their functional
form (\ref{ABCDform1}), the nonlinear coefficients are defined as before in (\ref{mu}), and
  \begin{equation}
     \omega_i'' = \frac
       {1 - \vg{i}^2 + \alpha_i^2 \delta^2 (c^2 - \vg{i}^2) + 4 \omega_i^2
       \alpha_i
       (\vg{i} - k / \omega_i)(\vg{i} - c^2 k / \omega_i)}
       {\omega_i (1 + \alpha_i^2 \delta^2)}.
     \elabel{omegakk}
  \end{equation}
We proceed further by integrating (\ref{ABCDeq2a})
   with respect to\ $\tau$ at constant $\eta_1$ from $0$ to $\tau_0$, and
   dividing by $\tau_0$. Letting $\tau_0 \rightarrow \infty$, and
   demanding that $A_2$ remains bounded in this limit, yields
   \begin{multline*}
     A_T = \frac{i}{2} \omega''_1 A_{\eta_1 \eta_1} + i \mu_1 |A|^2 A
         + i A \lim_{\tau_0 \rightarrow \infty} \left( \frac{1}{\tau_0}
           \int_{0}^{\tau_0} \left( 2 \mu_1 |B|^2 + \mu_2 |C|^2 + \mu_2 |D|^2
           \right) \, {\rm d} \tau \right) \\
          + i \mu_2 \lim_{\tau_0 \rightarrow \infty} \left( \frac{1}{\tau_0}
           \int_{0}^{\tau_0} B^* CD \, {\rm d} \tau \right), 
   \end{multline*}
where we have recalled that $A=A(\eta_1,T)$, and where the integrals are
taken at constant $\eta_1$. Noting the functional forms (\ref{ABCDform1}),
   we may use (\ref{etaxi}) to convert the terms involving $|B|^2$, $|C|^2$ and $|D|^2$ into integrals with respect to $\xi_1$, $\eta_2$ and $\xi_2$ respectively, still at fixed $\eta_1$, eventually yielding
   \begin{multline}
       A_T = \frac{i}{2} \omega''_1 A_{\eta_1 \eta_1} + i \mu_1 |A|^2 A 
          + i \mu_2 \lim_{\tau_0 \rightarrow \infty} \left( \frac{1}{\tau_0}
           \int_{0}^{\tau_0} B^* CD \, {\rm d} \tau \right) \\
         + i A \lim_{\tau_0 \rightarrow \infty} \frac{1}{\tau_0} \left(
           2 \mu_1 \int_{\eta_1}^{\eta_1+\tau_0} |B|^2 \, {\rm d} \xi_1
         +   \mu_2 \int_{\eta_1}^{\eta_1+\tau_0} |C|^2 \, {\rm d} \eta_2
         +   \mu_2 \int_{\eta_1}^{\eta_1+\tau_0} |D|^2 \, {\rm d} \xi_2
             \right).
      \elabel{Atemp} 
   \end{multline}
Here $\tau_0$ has been rescaled in each of the final three integrals, and to perform this rescaling we have assumed that $\vg{1}$ and $\vg{2}$ are both $O(1)$, with $(\vg{1}-\vg{2})=O(1)$.
Finally, we use the periodicity of $B$, $C$ and $D$ to convert the
   final three integral terms of (\ref{Atemp}) to averages over one period. Performing similar
   operations for (\ref{ABCDeq2b}), (\ref{ABCDeq2c}) and (\ref{ABCDeq2d}) gives
  \begin{equation}
     \begin{split}
       &A_T  = \frac{i}{2} \omega_1'' A_{\eta_1 \eta_1}
                  + i \mu_1 \left( |A|^2 + 2 \avg{B} \right) A
                  + i \mu_2 \left( \avg{C} + \avg{D} \right) A  
                  + i \mu_2 \lim_{\tau_0 \rightarrow \infty} \left( \frac{1}{\tau_0}
           \int_{0}^{\tau_0} B^* CD \, {\rm d} \tau \right), \\
       &B_T  =  -\frac{i}{2} \omega_1'' B_{\xi_1 \xi_1}
                  - i \mu_1 \left( 2 \avg{A} + |B|^2 \right) B
                  - i \mu_2 \left( \avg{C} + \avg{D} \right) B
                  - i \mu_2 \lim_{\tau_0 \rightarrow \infty} \left( \frac{1}{\tau_0}
           \int_{0}^{\tau_0} A^* CD \, {\rm d} \tau \right), \\
       &C_T  =  \frac{i}{2} \omega_2'' C_{\eta_2 \eta_2}
                  + i \mu_3 \left( |C|^2 + 2 \avg{D} \right) C
                  + i \mu_4 \left( \avg{A} + \avg{B} \right) C
                  + i \mu_4 \lim_{\tau_0 \rightarrow \infty} \left( \frac{1}{\tau_0}
           \int_{0}^{\tau_0} D^* AB \, {\rm d} \tau \right), \\
       &D_T  =  -\frac{i}{2} \omega_2'' D_{\xi_2 \xi_2}
                  - i \mu_3 \left( 2 \avg{C} + |D|^2 \right) D
                  - i \mu_4 \left( \avg{A} + \avg{B} \right) D
                  - i \mu_4 \lim_{\tau_0 \rightarrow \infty} \left( \frac{1}{\tau_0}
           \int_{0}^{\tau_0} C^* AB \, {\rm d} \tau \right) ,
     \end{split}
   \elabel{ABCDeq3}
  \end{equation}
where 
 \begin{align*}
   & \avg{A}(T) = \frac{1}{\lxlong} \int_0^{\lxlong} |A|^2 {\rm d} 
\eta_1, \quad
     \avg{B}(T) = \frac{1}{\lxlong} \int_0^{\lxlong} |B|^2 {\rm d} \xi_1, \\
   & \avg{C}(T) = \frac{1}{\lxlong} \int_0^{\lxlong} |C|^2 {\rm d} 
\eta_2, \quad
     \avg{D}(T) = \frac{1}{\lxlong} \int_0^{\lxlong} |D|^2 {\rm d} \xi_2.
\end{align*}
These {\nlcnls} equations govern the weakly nonlinear evolution of the wave
amplitudes in the general case, when $\vg{1}$, $\vg{2}$ and $(\vg{1} - \vg{2})$ are all $O(1)$.

If $\vg{1}$, $\vg{2}$ are $O(1)$ but $(\vg{1}-\vg{2})=O(\eps)$, then the averaging process described above fails, and we write
  \begin{equation}
   \vg{1} = \vgbar - \eps \vgdev, \quad
   \vg{2} = \vgbar + \eps \vgdev, \quad
   \mbox{where } \vgbar = \frac{\vg{1}+\vg{2}}{2}, \quad
   \vgdev = \frac{\vg{2} - \vg{1}}{2 \eps}.
   \elabel{vgbar}
\end{equation}
Then, rather than (\ref{ABCDform1}), the solvability conditions at $O(\eps)$ 
give $A=A(\overline \eta, T)$, $B=B(\overline \xi, T)$, $C=C(\overline 
\eta, T)$, $D=D(\overline \xi, T)$, where
  \begin{equation*}
   \etabar = \longx - \vgbar \tau, \quad
   \xibar = \longx + \vgbar \tau.
  \end{equation*}
At $O(\eps^3)$ we obtain equations similar to (\ref{ABCDeq2a})--(\ref{ABCDeq2d}), but with
  additional terms $-\vgdev A_\longx$, $\vgdev B_\longx$, $\vgdev C_\longx$,
  $-\vgdev D_\longx$ on the left-hand side. We then once again perform
  an averaging operation, yet this time $A$ and $C$ are now both
  functions of $\etabar$ and $T$, whilst $B$ and $D$ are both functions of
  $\xibar$ and $T$. Thus, the final integral terms can be considerably simplified, and rather than (\ref{ABCDeq3}) we obtain  
  \begin{equation}
     \begin{split}
       &A_T  - \vgdev A_\etabar =  \frac{i}{2} \omega_1'' A_{\etabar \etabar}
                  + i \mu_1 \left( |A|^2 + 2 \avg{B} \right) A
                  + i \mu_2 \left( |C|^2 + \avg{D} \right) A
                   + i \mu_2 \  \overline{B^*D}\  C, \\
       &B_T  + \vgdev B_\xibar =  -\frac{i}{2} \omega_1'' B_{\xibar \xibar}
                  - i \mu_1 \left( 2 \avg{A} + |B|^2 \right) B
                  - i \mu_2 \left( \avg{C} + |D|^2 \right) B
                   - i \mu_2 \  \overline{A^* C} \  D, \\
       &C_T  + \vgdev C_\etabar =  \frac{i}{2} \omega_2'' C_{\etabar \etabar}
                  + i \mu_3 \left( |C|^2 + 2 \avg{D} \right) C
                  + i \mu_4 \left( |A|^2 + \avg{B}  \right) C
                   +i \mu_4 \   \overline{D^* B} \  A,  \\
       &D_T  - \vgdev D_\xibar =  -\frac{i}{2} \omega_2'' D_{\xibar \xibar}
                  - i \mu_3 \left( 2 \avg{C} + |D|^2 \right) D
                  - i \mu_4 \left( \avg{A} + |B|^2 \right) D
                   -i\mu_4 \  \overline{C^* A} \   B,
     \end{split}
   \elabel{ABCDeq4}
  \end{equation}
with the coefficients defined as before. There is now a local coupling between the $A$ and $C$ modes, and between the $B$ and $D$ modes.   

If one permits $\vg{1}$ and $\vg{2}$ to be $O(\eps)$, then three other scenarios arise in which the averaging process fails. If $\vg{1}=O(\eps)$ but $\vg{2}=O(1)$, then using the obvious extension of the above analysis one obtains a system in which $A$ and $B$ are locally coupled, whilst $C$ and $D$ satisfy {\nlcnls} equations. If $\vg{2}=O(\eps)$ but $\vg{1}=O(1)$, then $C$ and $D$ are locally coupled, whilst $A$ and $B$ satisfy {\nlcnls} equations. If both $\vg{1}$ and $\vg{2}$ are both $O(\eps)$, then something slightly different occurs. Rather than (\ref{ABCDform1}), the solvability conditions at $O(\eps)$ give
$A=A(\longx,T), B=B(\longx,T), C=C(\longx,T), D=D(\longx,T)$.
At $O(\eps^3)$, the averaging operation immediately yields a system of {\cnls} equation for $A$, $B$, $C$ and $D$, with no non-local coupling, which is equivalent to system (\ref{ABCDeq5}) introduced below. However, none of these scenarios are further considered here, since $\vg{1}$ and $\vg{2}$ are typically $O(1)$ for the {\sg} system (\ref{1.2}).  
  
\subsection{The {\cnls} equations} \label{s:thirdway}

An alternative approach is to seek a system with linear
dispersive terms, unlike (\ref{ABCDeq1}), only local nonlinear coupling,
unlike (\ref{ABCDeq3}) and (\ref{ABCDeq4}), and with no a priori restrictions on the size
of the group velocities, unlike (\ref{ABCDeq3}), (\ref{ABCDeq4}) and the other systems mentioned at the end of Section~\ref{s:long}. 
Such a model is obtained, for instance, by allowing the wave
amplitudes $A$, $B$, $C$ and $D$ to depend on the long lengthscale
$\longx=\eps x$ and the slow timescale $\tau=\eps t$, whilst retaining
all terms up to order $O(\eps^3)$. The result is 
\begin{equation}
   \begin{split}
    A_{\tau}  + \vg{1} A_{\longx} - \frac{i}{2} \eps \omega''_1
    A_{\longx \longx} &
     = i \mu_1 \eps (|A|^2 + 2 |B|^2) A
     + i \mu_2 \eps (|C|^2 + |D|^2)A 
      + i\mu_2 \eps B^* C D, \\
    B_{\tau} - \vg{1} B_{\longx} + \frac{i}{2} \eps \omega''_1
    B_{\longx \longx } &
     = - i \mu_1 \eps (2 |A|^2 + |B|^2) B
       - i \mu_2 \eps  (|C|^2 + |D|^2)B
       - i \mu_2 \eps A^* C D, \\
    C_{\tau} + \vg{2} C_{\longx} - \frac{i}{2} \eps \omega''_2 
    C_{\longx \longx} &
     = i \mu_3 \eps (|C|^2 + 2 |D|^2) C
     + i \mu_4 \eps (|A|^2 + |B|^2)C
      + i \mu_4 \eps D^* A B, \\
    D_{\tau} - \vg{2} D_{\longx} + \frac{i}{2} \eps \omega''_2 
    D_{\longx \longx} &
     = - i \mu_3 \eps (2 |C|^2 + |D|^2) D
       - i \mu_4 \eps (|A|^2 + |B|^2)D
        -i \mu_4 \eps C^* AB.
   \end{split}
   \elabel{ABCDeq5}
  \end{equation}
In this form the linear dispersive and nonlinear terms are of the same
    relative order, but $O(\eps)$ smaller than the transport terms. If
    instead we had used the variables $X=\eps^2 x$, $T =\eps^2 t$,
then we would have had a balance between the transport terms and the nonlinear
terms, with the linear dispersive terms being relatively $O(\eps^2)$.
Only in certain special cases may one simply rescale $x$ and $t$ 
so that all terms of (\ref{ABCDeq5}) are simultaneously of the same
order. For instance, if only one wave is present, then in a reference
    frame moving at the group velocity of the wave one may use $T= \eps^2 t$ and $\chi = \eps x$. Similarly, if $B=D=0$, and
    $\vg{1}=\vg{2}$, one may perform the same operation to obtain
    coupled NLS equations for $A$ and $C$.  

Nevertheless, (\ref{ABCDeq5}) is an appealing model on purely physical
grounds. Higher order terms, both linear and nonlinear, could, in
principle, be dynamically important. Indeed, since the effects 
of transport, dispersion, and leading order nonlinearity are captured,
models of this type are widely used in studies of weakly nonlinear wave 
phenomena, as mentioned already in the Introduction. 
There is an analogy here with the well-known Boussinesq equations
often used to model the two-way propagation of water waves. Indeed,
any of the asymptotic models (\ref{ABCDeq1}), 
(\ref{ABCDeq3}) and (\ref{ABCDeq4}) already 
introduced, or those mentioned at the end of Section~\ref{s:long}, can be derived from (\ref{ABCDeq5}) by an {\ams} reduction on the appropriate length and time-scales.     
 
Note that one can deduce (\ref{ABCDeq5}) without recourse to a new 
asymptotic expansion. With the usual approach (e.g. 
\cite{Fauve,Thual}), the terms on the left-hand side of 
(\ref{ABCDeq5}) result from a Taylor series expansion of the 
dispersion relation about the dominant wavenumber of that packet. 
Only the first two terms are retained. The terms on the right-hand 
side are the relevant nonlinear terms, as already obtained in
(\ref{ABCDeq1}), for instance.


\section{Modulational instability of two-wave solutions} \label{s:2wave}

It is well known that dispersive weakly nonlinear systems of the type
being studied can support modulational instabilities. Some simple
examples of this for two-wave solutions were given in our previous
study \cite{Griffiths}. These results are briefly summarised in Section~\ref{s:previous}. Then, in Sections~\ref{s:new} and~\ref{s:cnls}, the analysis is extended to the other asymptotic models derived in Section~\ref{s:models}. 
 
\subsection{Plane waves} \label{s:plane2} 

We consider the modulational instability of spatially uniform coupled plane wave solutions of the form
  \begin{equation}
    A = A_0 \expf{i \Omega_a T},  \quad C = C_0 \expf{i \Omega_c T}, \quad B_0=D_0=0.
    \elabel{ABCDform2}
   \end{equation}
This is a solution of the {\nondis} (\ref{ABCDeq1}) model, the {\nlcnls} models
(\ref{ABCDeq3}) and (\ref{ABCDeq4}), and the {\cnls} model
(\ref{ABCDeq5}), provided
   \begin{equation*}
     \Omega_a = \mu_1 |A_0|^2 + \mu_2 |C_0|^2, \quad
     \Omega_c =  \mu_3 |C_0|^2 + \mu_4 |A_0|^2. 
   \end{equation*}
Substituting (\ref{ABCDform2}) into (\ref{uw1}) we obtain the leading order solution
in the form of two counter-propagating waves:
   \begin{equation}
    \begin{pmatrix} u \\ w \end{pmatrix} = \eps A_0 \expf{i(k x - 
(\omega_1 - \eps^2 \Omega_a) t)}
    \begin{pmatrix} 1 \\ \alpha_1 \end{pmatrix}
    + \eps C_0 \expf{i(k x - (\omega_2 -\eps^2 \Omega_c) t)}
    \begin{pmatrix} 1 \\ \alpha_2 \end{pmatrix} + {\rm c.c.} + O(\eps^2). 
    \elabel{uw2waves}
   \end{equation}
This solution describes a weakly nonlinear interaction 
between the two components. For $\eps \neq 0$, it generalises
   the linear energy exchange solutions of Section~\ref{s:linear} by 
nonlinearity-induced corrections
   of $O(\eps^2)$ to the frequencies $\omega_{1,2}(k)$ of the linear
   waves, given by (\ref{omega12}). For $\eps \to 0$, it reduces to 
the two-wave linear solution (\ref{2.3}) if
   \begin{equation}
    \eps A_0 = \frac{|\alpha_2| \amp}{2(\alpha_1 + |\alpha_2|)}, \;\;\;
    \eps C_0 = \frac{\alpha_1 \amp}{2(\alpha_1 + |\alpha_2|)}. 
   \elabel{ACvals}
   \end{equation}
In the case $c=1$, (\ref{uw2waves}) matches the weakly nonlinear expansion
(\ref{3.2}) of the exact two-wave energy-exchange solution
(\ref{3.1}). Since $\alpha_1=1$ and $\alpha_2=-\delta^{-2}$ at $c=1$, from (\ref{mu}) 
$\mu_1 = \mu_2 = \mu_4 = 0$, and $\mu_3 = (1+\delta^2)^3 / (4 
\omega_2 \delta^4)$, so that taking
  \begin{equation*}
   \eps A_0 = \frac{\kappa}{1+\delta^2}, \quad
   \eps C_0 = \frac{\kappa \delta^2}{1+\delta^2}, \quad
   B_0 = D_0 = 0,
  \end{equation*}
we find $\Omega_a = 0$, $\eps^2 \Omega_c  = \mu_3
   |\eps C_0|^2 = \kappa^2 (1+\delta^2) / 4 \omega_2$, in agreement
with (\ref{3.2}).

\subsection{Previous results} \label{s:previous}

In~\cite{Griffiths} we considered the stability of the coupled plane wave solutions (\ref{ABCDform2}) in the context of two of the asymptotic models of Section~\ref{s:models}. Firstly, we showed that the {\nondis} equations (\ref{ABCDeq1}) with $B=D=0$ do not support any modulational instabilities. Secondly, we studied instabilities with modulations on the long lengthscale $\longx=\eps x$, with $\vg{1}-\vg{2}$ non-zero and $O(1)$, so that we used (\ref{ABCDeq3}) with $B=D=0$. However, the dynamics of that system are relatively simple, since as in \cite{Gibbon} the equations may be transformed to two uncoupled NLS equations using the nonlocal transformation
  \begin{equation}
    \begin{split}
      A(\eta_1, T) & = \tilde A(\eta_1, T)
        \exp \left( i \mu_2 \int_0^T \avg{C}(T') {\rm d} T' \right), \\ 
      C(\eta_2, T) & = \tilde C(\eta_2, T)
        \exp \left( i \mu_4 \int_0^T \avg{A}(T') {\rm d} T' \right).
    \end{split}
   \elabel{trans}
  \end{equation}
Alternatively, stability results can be established directly by looking for solutions of (\ref{ABCDeq3}) of the form 
  \begin{equation}
  \begin{split} 
   A(\eta_1,T) & = A_0(T) \left( 1 + a_1 \expf{i \left( \kdl \eta_1 - \omegadss_a T \right)}
                 + a_2^* \expf{-i \left( \kdl \eta_1 - \omegadss_a^* T \right) } \right), \\
   C(\eta_2,T) & =  C_0(T) \left( 1 + c_1 \expf{i \left( \kdl \eta_2 - \omegadss_c T \right)}
                 + c_2^* \expf{-i \left( \kdl \eta_2 - \omegadss_c^* T \right) } \right),
   \end{split}
   \elabel{ACform} 
  \end{equation}
representing sinusoidal perturbations to the {\cw} solution (\ref{ABCDform2}). Neglecting terms quadratic in disturbance amplitude, (\ref{ABCDeq3}) leads to two uncoupled linear systems, as expected from (\ref{trans}), with corresponding dispersion relations 
 \begin{equation}
        \omegadss_a^2 = - \mu_1 |A_0|^2 \omega_1''  \kdl^2
             + \frac{{\omega_1''}^2 \kdl^4}{4}, \quad
        \omegadss_c^2 = - \mu_3 |C_0|^2 \omega_2''  \kdl^2
             + \frac{{\omega_2''}^2 \kdl^4}{4}.
  \elabel{rates1}
 \end{equation}
Using (\ref{scales}) and (\ref{etaxi})  we can rewrite these in terms of an unscaled disturbance wavenumber $\kd$ and frequency $\omegad$, via 
 \begin{equation}
  \eps \kdl=\kd, \quad
  \eps^2 \omegadss_a=\omegad_a-\kd \vg{1}, \quad
  \eps^2 \omegadss_c=\omegad_c-\kd \vg{2}, 
  \elabel{dscales} 
 \end{equation} 
 giving 
   \begin{equation}
      \left( \omegad_a - \kd \vg{1} \right)^2 =
      - \mu_1 |\eps A_0|^2 \omega_1'' \kd^2 + \frac{{\omega_1''}^2 \kd^4}{4}, \quad
     \left( \omegad_c - \kd \vg{2} \right)^2 = 
      - \mu_3 |\eps C_0|^2 \omega_2'' \kd^2 + \frac{{\omega_2''}^2 \kd^4}{4},
     \elabel{rates2}
    \end{equation}
where for formal validity we need $A_0, C_0 = O(1)$, $\kd =O(\eps)$, and $(\vg{1}-\vg{2})=O(1)$. 

From (\ref{mu}), we see that $\mu_1>0$ and $\mu_3>0$, so there
will be instability only if $\omega_1''>0$ or $\omega_2''>0$.
Using (\ref{omegakk}), we can show that $\omega_2''$ is always positive.
Thus for sufficiently small $\kd$ there will be a root for $\omegad_c$ with positive imaginary part, corresponding to an unstable mode. Using (\ref{omega12}), we can show that $\omega_1''<0$ for small
$k$, whilst $\omega_1''>0$ for large $k$,  tending to zero at infinity.
Thus $\omegad_a$ is real for small $k$, and hence does not correspond to
an unstable mode, whilst for larger $k$, where $\omega_1''>0$,
it will correspond to an unstable mode for sufficiently small $\kd$.

\subsection{New results: the {\cnls} equations} \label{s:new} 

We now consider instability in the context of the two other asymptotic models of Section~\ref{s:models}. The first model is for modulations on the long lengthscale $\longx=\eps x$, but with $(\vg{1}-\vg{2}) = 
O(\eps)$. Then we must use equations (\ref{ABCDeq4}), with $B=D=0$, giving a system that is locally coupled in $A$ and $C$. The second model is the {\cnls} equations (\ref{ABCDeq5}), with $B=D=0$. However, with the appropriate rescaling, it is easy to see that these two systems of equations are equivalent. Therefore, even though the two models were derived in different ways, they may be analysed together.     

We look for disturbances to the coupled-wave solution (\ref{ABCDform2})  in the form
   \begin{align*}
     A & = A_0(T) \left( 1 + a_1 \expf{i(\kd x - \omegad_{ac} t)}
                + a_2^* \expf{-i(\kd x - \omegad_{ac}^* t)} \right), \\
     C & = C_0(T) \left( 1 + c_1 \expf{i(\kd x - \omegad_{ac} t)}
                + c_2^* \expf{-i(\kd x - \omegad_{ac}^* t)} \right),
   \end{align*}
where we find it convenient to revert to the original unscaled variables $x$, $t$ and work in terms of 
the unscaled disturbance frequency $\omegad$ and wavenumber $\kd$. 
Substituting into (\ref{ABCDeq4}) or (\ref{ABCDeq5}) we obtain:
   \begin{equation*}
    \begin{split}
     \left( \omegad_{ac} - \kd \vg{1} - \frac{1}{2} \omega''_1 \kd^2 + \mu_1
     |\eps A_0|^2 \right) a_1 + \mu_1 |\eps A_0|^2 a_2 + \mu_2 |\eps
C_0|^2 c_1 + \mu_2 |\eps C_0|^2 c_2 & = 0, \\
     - \mu_1 |\eps A_0|^2 a_1 + \left( \omegad_{ac} - \kd \vg{1} +
\frac{1}{2} \omega''_1 \kd^2 - \mu_1 |\eps A_0|^2 \right) a_2 -
\mu_2 |\eps C_0|^2 c_1 -\mu_2 |\eps C_0|^2 c_2 & = 0, \\
     \mu_4 |\eps A_0|^2 a_1 + \mu_4 |\eps A_0|^2 a_2 + \left( \omegad_{ac} 
- \kd \vg{2} -
     \frac{1}{2} \omega''_2 \kd^2 + \mu_3 |\eps C_0|^2 \right) c_1
     + \mu_3 |\eps C_0|^2 c_2 & = 0, \\
     -\mu_4 |\eps A_0|^2 a_1 - \mu_4 |\eps A_0|^2 a_2 - \mu_3 |\eps 
C_0|^2 c_1 + \left( \omegad_{ac} - \kd \vg{2} + \frac{1}{2}
\omega''_2 \kd^2 - \mu_3 |\eps C_0|^2 \right) c_2 & = 0.
    \end{split}
   \end{equation*}
This four-by-four linear homogeneous
system of equations must have a vanishing determinant, from which we find that
   \begin{equation}
     \qa(\ps,\kd) \qc(\ps,\kd) = R, \quad  \mbox{where }
     R  = \mu_2 \mu_4 \omega''_1 \omega''_2 |\eps A_0|^2 |\eps 
C_0|^2, \elabel{sw1} 
 \end{equation}
 \begin{equation} 
     \qa(\ps,\kd) = \left( \ps - \vg{1} \right)^2
            - \frac{{\omega''_1}^2 \kd^2}{4}
           + \mu_1 \omega''_1 |\eps A_0|^2 , \quad
     \qc(\ps,\kd) = \left( \ps - \vg{2} \right)^2
           - \frac{{\omega''_2}^2 \kd^2}{4}
           + \mu_3 \omega''_2 |\eps C_0|^2.  
       \elabel{sw2}
   \end{equation}
Here $\ps=\omegad_{ac}/\kd$ is the phase speed of the disturbance, whilst $R$ is proportional to the coupling constants in (\ref{ABCDeq4}) and (\ref{ABCDeq5}). Note that $\qa=0$ and $\qc=0$ are the dispersion relations for the uncoupled NLS equations for $A$ and $C$ respectively, cf.\ (\ref{rates2}). Equation (\ref{sw1}) is a fourth-order dispersion relation with real coefficients for $\ps$. Complex solutions for $\ps$ imply modulational instability.

\subsection{Instability analysis: the {\cnls} equations} \label{s:cnls}

Working within the context of (\ref{ABCDeq4}), for formal validity of (\ref{sw1}) we need $A_0, C_0 = O(1)$, $\kd =O(\eps)$, and $(\vg{1}-\vg{2})=O(\eps)$. However, within the context of (\ref{ABCDeq5}), there are no specific scale restrictions, and we may investigate different scaling regimes.    

We first probe the effects of the dispersive terms for modulations on
the {\superlong} lengthscale, by taking $\kd = \eps^2 \kdsl$, with
$\kdsl=O(1)$ and $\eps \ll 1$. Writing $\ps=\ps_1 + \eps^2 \ps_2 +
\cdots $, one may derive expressions for the four roots as
  \begin{align*}
   (\ps - \vg{1} )^2 & = - \eps^2 \mu_1 \omega''_1 |A_0|^2
     + \eps^4 \left( \frac{1}{4} {\omega''_1}^2 \kdsl^2
     + \frac{\mu_2 \mu_4 \omega''_1 \omega''_2 |A_0|^2 |C_0|^2 }
       {(\vg{1}-\vg{2})^2} \right) + O(\eps^6), \\
   (\ps - \vg{2} )^2 & = - \eps^2 \mu_3 \omega''_2 |C_0|^2
     + \eps^4 \left( \frac{1}{4} {\omega''_2}^2 \kdsl^2
     + \frac{\mu_2 \mu_4 \omega''_1 \omega''_2 |A_0|^2 |C_0|^2}
       {(\vg{1}-\vg{2})^2} \right) + O(\eps^6).
   \end{align*}
Since $\mu_3>0$ and $\omega''_2>0$, at least one pair of roots for
$\ps$ is complex, and hence there is always an unstable mode. Since $\mu_1>0$,
there will be an additional unstable mode if $\omega''_1>0$.
Thus, our analysis suggests that the dispersive terms
are destabilising for disturbances on the {\superlong} lengthscale,
leading to an instability with a growth rate of $O(\eps^3)$. This is
consistent with the nondispersive analysis of~\cite{Griffiths}, mentioned in Section~\ref{s:previous}, which showed there will be no instabilities on the {\superlong} lengthscale with growth rates of $O(\eps^2)$. 

We can recover results for modulations on the long lengthscale by taking $\kd =
\eps \kdl$, with $\kdl=O(1)$ and $\eps \ll 1$. If
$(\vg{1}-\vg{2})=O(1)$, then writing $\ps = \ps_1 + \eps \ps_2 +
\cdots$, one may derive expressions for the four roots as 
  \begin{equation}
   \begin{split}
    \left( \ps - \vg{1} \right)^2 & = -\eps^2 \left( \mu_1 \omega''_1 
|A_0|^2 - \frac{1}{4} {\omega''_1}^2 \kdl^2 \right) + \frac{\eps^4 
\mu_2 \mu_4 \omega''_1 \omega''_2 |A_0|^2 
|C_0|^2}{(\vg{1}-\vg{2})^2} + O(\eps^6), \\
    \left( \ps - \vg{2} \right)^2 & = -\eps^2 \left( \mu_3 \omega''_2
|C_0|^2 - \frac{1}{4} {\omega''_2}^2 \kdl^2 \right) + \frac{\eps^4
\mu_2 \mu_4 \omega''_1 \omega''_2 |A_0|^2
|C_0|^2}{(\vg{1}-\vg{2})^2} + O(\eps^6). 
    \end{split}
   \elabel{vlong}
  \end{equation}
Since $\mu_3>0$ and $\omega''_2>0$, there will be an instability with
a growth rate of $O(\eps^2)$ if $\kdl^2 < 4 \mu_3 |C_0|^2 /
\omega''_2$. Since $\mu_1>0$, there will be an additional unstable
mode if $\omega''_1>0$ and $\kdl^2 < 4 \mu_1 |A_0|^2 / \omega''_1$.
This is consistent with (\ref{rates1}) and (\ref{rates2}), derived 
formally from the {\nlcnls} equations. 
If $(\vg{1}-\vg{2})=O(\eps)$, then no further approximation of (\ref{sw1}) is possible, 
since it is already the asymptotically consistent model for this regime.

\begin{figure}
\begin{center}
\begin{picture}(155,45)(0,0)
\put(-7,-3){\includegraphics[width=160mm]{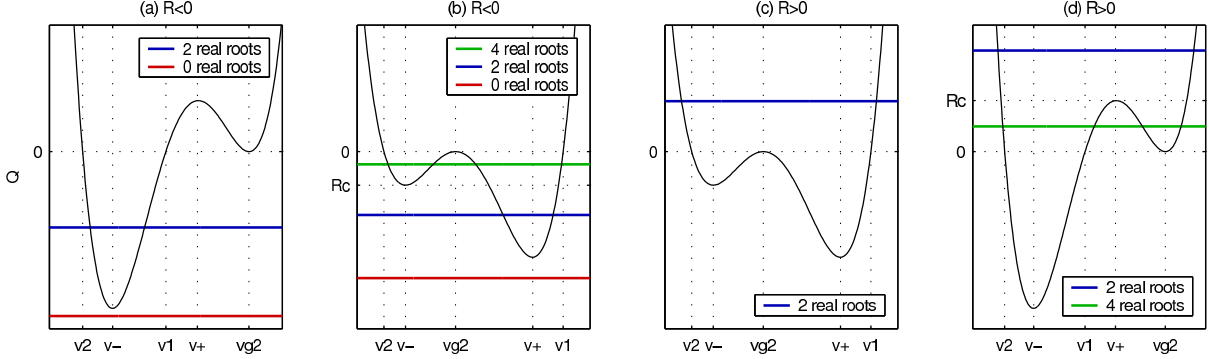}}
\end{picture}
\end{center}
\caption{{\small Schematic representations of the possible stability scenarios at $\kd=\kcc$. Plotted in black in each case is the coupled dispersion function $Q=\qa(\ps,\kcc) \qc(\ps,\kcc)$ as a function of $\ps$, as defined in (\ref{sw1}). The horizontal lines denote values of $R$ with differing stability properties.}} 
\label{f:qv}
\end{figure}

More generally, we now consider the properties of the roots $\ps(\kd)$
of (\ref{sw1}) without any scale restrictions on $\kd$ or $\eps$. First,
if $\kd \gg 1$, then the roots of (\ref{sw1}) are approximately $\ps = \pm \omega''_1 \kd / 2$ and $\ps = \pm \omega''_2 \kd / 2$, so that there is stability. We then consider how the roots vary as
$\kd$ is reduced from infinity, with all other parameters fixed. Note
that these other parameters consist of the plane wave parameters $k$,
$|\eps A_0|$ and $|\eps C_0|$, and the system parameters $c$ and $\delta$.
We consider the graph of $Q=\qa(\ps,\kd)\qc(\ps,\kd)$ as a 
fourth order polynomial in $\ps$, as $\kd$ is varied. Stability occurs
whenever this graph intersects the line $Q=R$ four times,
and otherwise there is instability. Note that $R$ is independent
of $\kd$ and so this line is immovable as $\kd$ is varied.
Noting that $\mu_2$, $\mu_4$ and $\omega''_2$ are all positive, $R$
takes the same sign as $\omega''_1$. There are thus two possible
cases to consider:
   \begin{enumerate}
   \item Suppose first that $\omega''_1 < 0$, so that $R < 0$.
   For sufficiently large $\kd$, both uncoupled modes are stable
   and so $\qa\qc$ has four real zeros, $\ps_{i}, i=1,2,3,4$ where
   $\ps_1 > \ps_2, \ps_3 > \ps_4$ are the zeros of $\qa,\qc$
   respectively. Since $\mu_1>0$ and $\omega''_1 < 0$ here,
   $\ps_{1,2}$ are real for all $\kd$, from (\ref{sw2}). But as $\kd$
   is reduced from infinity, there is a critical value of
   $\kd=\kcc=2|\eps C_0|(\mu_3/\omega''_2)^{1/2}$, at which
   $\ps_3=\ps_4=\vg{2}$, and below which $\ps_{3,4}$ become complex,
   so that the uncoupled $C$-mode is then unstable. The issue then is
   whether the coupling in (\ref{sw1}) suppresses or enhances this instability.
   Consider then the situation when $\kd=\kcc$, so that
   $\qc = (\ps - \vg{2})^2$ and $\qa = (\ps-\vg{1})^2 - \beta^2$, where
   \begin{equation}
    \beta^2 = (\omega''_1 / \omega''_2)
            \left( \mu_3 \omega''_1 |\eps C_0|^2
                 - \mu_1 \omega''_2 |\eps A_0|^2 \right) > 0.
    \elabel{betasq}
   \end{equation}
   There are two sub-cases to consider: 
   \begin{enumerate}
   \item If $|\vg{2}-\vg{1}| \ge \beta$, then the double zero $\vg{2}$ of
   $\qc(\ps,\kcc)$ does not lie in $(\ps_2,\ps_1)$. This scenario is depicted in Figure~\ref{f:qv}a.
   Since $R<0$, $\qa(\ps,\kcc) \qc(\ps,\kcc)=R$ must have at least
   two complex roots, so that the coupled system
   is unstable at $\kd=\kcc$, and therefore for some $\kd >
   \kcc$. In this sense we can say that coupling is destabilizing. 
   Indeed, in this case the system remains unstable for
   all $\kd < \kd_c$, where $\kd_c > \kcc$ is the critical value at
   which the system first becomes unstable, and is the value of $\kd$
   at which the line $Q=R$ is tangent to the graph of $Q=\qa \qc$.
   \item If $|\vg{2}-\vg{1}| < \beta$, then the double
   zero $\vg{2}$ of $\qc(\ps,\kcc)$ lies in $(\ps_2,\ps_1)$.  This scenario is depicted in Figure~\ref{f:qv}b.  
   Stability at $\kd=\kcc$ is then determined by  
    \begin{equation*}
     R_c={\rm max}(\qa(\ps_\pm,\kcc) \qc(\ps_\pm,\kcc) ), \quad
     \ps_\pm = \vg{2} +
       \frac{ 3 (\vg{1}-\vg{2}) \pm \sqrt{(\vg{1}-\vg{2})^2 + 8 \beta^2}}{4},
    \end{equation*}
   where $\ps_{\pm}$ are two of the turning points of $Q=\qa(\ps,\kcc)\qc(\ps,\kcc)$, 
   the third being at $\vg{2}$.
   When $R < R_c$, $\qa(\ps,\kcc) \qc(\ps,\kcc)=R$ has at least two complex roots, 
   and we can say that coupling is destabilizing, whereas if $R_c <  R < 0$ there 
   are four real roots of $\qa(\ps,\kcc)\qc(\ps,\kcc)=R$, and we can say that coupling is stabilizing.   
   But here, in both scenarios, the stability situation may change
   again as $\kd$ is further reduced below $\kcc$.
   \end{enumerate}
   \item Suppose next that $\omega''_1 > 0$, so that $R>0$. For
sufficiently large $\kd$, both uncoupled modes are stable, whilst for
sufficiently small $\kd$, both uncoupled modes are unstable. Suppose
that as $\kd$ is reduced from infinity, one of the uncoupled modes
becomes unstable for some value of $\kd$, while the other, for this
value of $\kd$ is still stable. To be explicit, let us suppose that it
is the $C$-mode which becomes unstable, so that $\kd=\kcc$ as above.
The other case can be treated in the same way.
Now at $\kd=\kcc$ our hypothesis that the $A$-mode is stable again
implies that $\beta^2>0$, and once again there are two sub-cases to consider:
\begin{enumerate}
\item If $|\vg{2}-\vg{1}| \le \beta$, then the double zero $\vg{2}$ of $\qc(\ps,\kcc)$
lies in $[\ps_2,\ps_1]$. This scenario is depicted in Figure~\ref{f:qv}c. 
Since $R > 0$ and all turning points of $Q=\qa(\ps,\kcc) \qc(\ps,\kcc)$ lie in $Q \le 0$, the coupled system is unstable at $\kd=\kcc$, and therefore for
some $\kd > \kcc$, and in this sense we can say that coupling is
destabilizing. Indeed, in this case the system remains unstable for
all $\kd<\kd_c$, where $\kd_c>\kcc$ is the critical value at which
the system first becomes unstable, and is the value of $\kd$ at which
the line $Q=R$ is tangent to the graph of $Q=\qa\qc$.
\item If $|\vg{2}-\vg{1}| > \beta$, then the double zero
$\vg{2}$ of $\qc(\ps,\kcc)$ does not lie in $[\ps_2,\ps_1]$. This scenario is depicted in Figure~\ref{f:qv}d. 
Stability at $\kd=\kcc$ is then once again determined by $R_c$, defined as before, which is now the value of the single turning point with $\qa\qc>0$. When $R>R_c$, $\qa(\ps,\kcc)\qc(\ps,\kcc)=R$ has two complex roots so coupling is destabilizing, whereas if $0 < R < R_c$ there are four real roots of $\qa(\ps,\kcc)\qc(\ps,\kcc)=R$ and coupling is stabilizing. But here, in both scenarios, the stability situation may change again as $\kd$ is further reduced below $\kcc$.
\end{enumerate}
\end{enumerate}

\subsection{Summary}

We have considered how modulational instabilities of spatially uniform coupled two-wave solutions are predicted by the four weakly nonlinear models of Section~\ref{s:models}. As noted in Section~\ref{s:previous}, the 
{\nondis} equations (\ref{ABCDeq1}) give an unambiguous prediction of 
stability for disturbances with $\kd$ of $O(\eps^2)$ on a timescale $\sim \eps^{-2}$. Further, the {\nlcnls} equations 
(\ref{ABCDeq3}) give an unambiguous prediction 
of instability for disturbances with $\kd$ of $O(\eps)$
on a timescale $\sim \eps^{-2}$, 
whenever $\vg{1}-\vg{2}$ is $O(1)$, and either 
$\omega''_1 > 0$ or $\omega''_2>0$ (the latter is always true here).
As shown in Sections~\ref{s:new} and~\ref{s:cnls}, the {\cnls} equations (\ref{ABCDeq4}) and (\ref{ABCDeq5}) give a dispersion relation which is harder to analyse, but we showed that it reproduces the results 
for disturbances with $\kd$ of $O(\eps)$ or $O(\eps^2)$ in their respective
regimes of validity.  
However, in the context of (\ref{ABCDeq5}), the {\cnls} dispersion relation also gives predictions which are strictly outside its asymptotic range of validity. Such predictions need to be tested independently, for instance, 
by numerical simulations of the full unapproximated {\sg} equations. 
We perform such tests in Section~\ref{s:numerics}, by numerically
determining the behaviour of system (\ref{1.2}).


\section{Modulational instability of four-wave solutions} \label{s:4wave}

We now discuss the modulational instability of four-wave solutions. We use the same ideas as in Section~\ref{s:2wave}, except that we restrict our analysis to the {\nlcnls} equations, derived  for modulations on the long lengthscale $\longx = \eps x$. 

\subsection{Plane waves} \label{s:plane}

We start by examining the form of spatially uniform coupled plane waves. For such solutions,  the 
{\nlcnls} models (\ref{ABCDeq3}) and (\ref{ABCDeq4}) reduce to the same system. Looking for solutions of the form
   \begin{equation}
    A = A_0 \expf{i \Omega_{a} T},  \quad B = B_0 \expf{-i \Omega_{b} T}, \quad
    C = C_0 \expf{i \Omega_{c} T}, \quad  D = D_0 \expf{-i \Omega_{d} T},
    \elabel{ABCDform3}
   \end{equation}
we see that a balance is only possible if
 \begin{equation}
   \Omega_a-\Omega_b=\Omega_c-\Omega_d. 
   \elabel{4balance1}
 \end{equation}
Substitution of (\ref{ABCDform3}) into (\ref{ABCDeq3}) or (\ref{ABCDeq4}) gives
   \begin{align*}
    \Omega_a & = \mu_1(|A_0|^2 + 2 |B_0|^2) + \mu_2 (|C_0|^2 + |D_0|^2) 
                               +  \mu_2 \amps^* / |A_0|^2, \\
    \Omega_b & = \mu_1(2|A_0|^2 + |B_0|^2)  + \mu_2 (|C_0|^2 + |D_0|^2)
                               + \mu_2 \amps^* / |B_0|^2, \\
    \Omega_c & =  \mu_3(|C_0|^2 +  2|D_0|^2) + \mu_4 (|A_0|^2 + |B_0|^2)
                               + \mu_4 \amps / |C_0|^2 ,\\
    \Omega_d & = \mu_3 (2|C_0|^2 + |D_0|^2) + \mu_4(|A_0|^2 + |B_0|^2)
                               + \mu_4 \amps / |D_0|^2,
   \end{align*}
 where
  \begin{equation}
    \amps= A_0 B_0 C_0^* D_0^*.
    \elabel{amps}
  \end{equation}
Thus, for solutions of the form (\ref{ABCDform3}), the condition (\ref{4balance1}) for balance  becomes
 \begin{equation}
   \left(  |A_0|^2 - |B_0|^2 \right) \left( \mu_1 + \frac{\mu_2 C_0 D_0}{A_0 B_0} \right) =
   \left( |C_0|^2 - |D_0|^2 \right) \left( \mu_3 + \frac{\mu_4 A_0 B_0}{C_0 D_0} \right). 
    \label{4balance2}
 \end{equation}  
If (\ref{4balance2}) is satisfied, then we may substitute (\ref{ABCDform3}) into (\ref{uw1}) to obtain a solution
in the form of two pairs of counter-propagating waves,
   \begin{multline}
    \begin{pmatrix} u \\ w \end{pmatrix} = \eps \left( A_0 \expf{i(k x - 
(\omega_1 - \eps^2 \Omega_a ) t)} + B_0
    \expf{i(k x + ( \omega_1 - \eps^2 \Omega_b ) t)}\right )
    \begin{pmatrix} 1 \\ \alpha_1 \end{pmatrix} \\
    + \eps \left( C_0 \expf{i(k x - (\omega_2 - \eps^2 \Omega_c ) t)}
    + D_0 \expf{i(k x + ( \omega_2 - \eps^2 \Omega_d ) t)}\right )
    \begin{pmatrix} 1 \\ \alpha_2 \end{pmatrix} + {\rm c.c.} + O(\eps^2) .
    \elabel{uw3}
   \end{multline}
This solution generalises the linear energy exchange solutions of Section~\ref{s:linear} by 
nonlinearly-induced corrections
   of $O(\eps^2)$ to the frequencies $\omega_{1,2}(k)$ of the linear
   waves, given by (\ref{omega12}). For $\eps \to 0$, it reduces to 
   the four-wave linear solution (\ref{2.6}) if
   \begin{equation}
    \eps A_0 = \eps B_0
      = \frac{|\alpha_2| \amp}{4(\alpha_1 + |\alpha_2|)}, \;\;\;\;
    \eps C_0 = \eps D_0
      = \frac{\alpha_1 \amp}{4(\alpha_1 + |\alpha_2|)}, 
    \elabel{ABCDvals}
   \end{equation}
which clearly satisfies the balance condition (\ref{4balance2}). 

\subsection{The {\nlcnls} equations: $(\vg{1}-\vg{2})=O(1)$}

We first study the instability of the plane-wave solution (\ref{ABCDform3}) in the case with $\vg{1}$ and $\vg{2}$ both $O(1)$, and $(\vg{1}-\vg{2})$ non-zero and $O(1)$. Accordingly we use equations (\ref{ABCDeq3}), and we look for solutions in the form (\ref{ACform}), with corresponding forms for $B$ and $D$. Neglecting terms quadratic in disturbance amplitude, this yields four linear systems of the form
 \begin{align*}
   \left( \omegadss_a - \frac{1}{2} \omega''_1 \kdl^2 + \mu_1
     |A_0|^2 - \frac{\mu_2 \amps^*}{|A_0|^2} \right) a_1 + \mu_1 |A_0|^2 a_2 & = 0, \\
  - \mu_1 |A_0|^2 a_1 + \left( \omegadss_a + \frac{1}{2} \omega''_1 \kdl^2 
  - \mu_1 |A_0|^2 + \frac{\mu_2 \amps}{|A_0|^2} \right) a_2 & = 0.
  \end{align*}
Note that although this system might appear only to involve $|A_0|$, 
in fact the terms involving $\amps$ give a coupling to the other basic state waves via $B_0$, $C_0$ and $D_0$, unlike the two-wave case.
Each system has an associated dispersion relation of the form
\begin{equation*}
    \omegadss_a^2 + \frac{\mu_2}{|A_0|^2} \left( \amps-\amps^* \right) \omegadss_a
   - \left( \frac{1}{2} \omega''_1 \kdl^2 + \frac{\mu_2 \amps}{|A_0|^2} - \mu_1 |A_0|^2 \right) 
      \left( \frac{1}{2} \omega''_1 \kdl^2 + \frac{\mu_2 \amps^*}{|A_0|^2} - \mu_1 |A_0|^2 \right) 
   + \mu_1^2 |A_0|^4 = 0.
 \end{equation*}
When $A_0$, $B_0$, $C_0$ and $D_0$ all have the same phase, as is the case for (\ref{ABCDvals}) for instance, then $\amps = |A_0 B_0 C_0 D_0|$, and the dispersion relation reduces to
 \begin{equation}
  \omegadss_a^2 = \left( \frac{1}{2} \omega''_1 \kdl^2 -  \mu_1 |A_0|^2 + \frac{\mu_2 |B_0 C_0  D_0|}{|A_0|}  \right)^2 - \mu_1^2 |A_0|^4. 
  \elabel{rates3}
 \end{equation}
Reverting to dimensional variables using (\ref{dscales}) gives 
\begin{equation}
   \left( \omegad_a - \kd \vg{1} \right)^2 = 
     \left( \frac{1}{2} \omega''_1 \kd^2 -  (1-r_a) \mu_1 |\eps A_0|^2\right)^2 - \mu_1^2 |\eps A_0|^4, 
     \; \; \; r_a = \frac{\mu_2}{\mu_1} \left| \frac{B_0 C_0 D_0}{A_0^3} \right|, 
  \elabel{rates4a}
 \end{equation}
with corresponding forms for $\omegad_b$, $\omegad_c$ and $\omegad_d$.  
Although equations (\ref{rates3}) and (\ref{rates4a}) have a similar form to their two-wave counterparts (\ref{rates1}) and (\ref{rates2}), there are important differences. For instance, consider the instability of the $A$-mode, as given by (\ref{rates4a}). 

When $\omega''_1>0$, (\ref{rates2}) shows that for the two-wave solution a maximum growth rate of $s_*=\mu_1 |\eps A_0|^2$ is achieved by the $A$-mode when $\kd^2=\kd_*^2=2\mu_1 |\eps A_0|^2/|\omega''_1|>0$. However, for the four-wave solution, instability of the $A$-mode is instead determined by the parameter $r_a$, and only if $r_a=0$ are the two-wave results recovered. If $0 < r_a \le 1$, then (\ref{rates4a}) is minimised at $\kd^2=\kd_*^2 (1 - r_a)$, so that the disturbance wavenumber of the most unstable $A$-mode is reduced, although the growth rate remains $s_*$. If $r_a > 1$, (\ref{rates4a}) is instead minimised at $\kd=0$, with value $s_*^2(r_a^2-2r_a)>-s_*^2$. Therefore, if $1<r_a<2$ the most unstable $A$-mode occurs at $\kd=0$ with a reduced growth rate $s<s_*$, whilst if $r_a \ge 2$ the $A$-mode is stable. Thus, a two-wave solution can yield an instability in the $A$-mode whilst a corresponding four-wave solution does not.   

When $\omega''_1<0$, (\ref{rates2}) shows that the two-wave solution does not lead to an instability in the $A$-mode. However, when $r_a \ge 1$, (\ref{rates4a}) is minimised at $\kd^2=(r_a-1)\kd_*^2$, leading to an instability in the $A$-mode with growth rate $s_*$. If $0 < r_a < 1$, (\ref{rates4a}) is instead minimised at $\kd=0$, with value $s_*^2 (r_a^2 - 2 r_a) > -s_*^2$, leading to an instability in the $A$-mode with a reduced growth rate $s<s_*$. Thus, a four-wave solution can yield an instability in the $A$-mode whilst a corresponding two-wave solution does not.  

There are corresponding conclusions for the $B$ mode, and for the $C$ and $D$-modes (for which $\omega''_2>0$), and the overall stability of (\ref{ABCDform3}) will be determined by the most unstable of these four modes.  
Although the possible stabilising and destabilising effects outlined above are of interest, we do not consider the details any further here. 


\subsection{The {\nlcnls} equations: $(\vg{1}-\vg{2})=O(\eps)$} \label{s:nlcnls4}

We now turn to the case with $\vg{1}$ and $\vg{2}$ both $O(1)$, but $(\vg{1}-\vg{2})=O(\eps)$. Accordingly we use equations (\ref{ABCDeq4}), and we look for solutions in the form
 \begin{align*}
   A(\etabar,T) & = A_0(T) (1 + a_1 \expf{i (\kdl \etabar - \omegadss_{ac} T)}
                 + a_2^* \expf{-i (\kdl \etabar - \omegadss_{ac}^* T) } ), \\
   C(\etabar,T) & =  C_0(T) (1 + c_1 \expf{i (\kdl \etabar - \omegadss_{ac} T)}
                 + c_2^* \expf{-i (\kdl \etabar - \omegadss_{ac}^* T) } ), 
 \end{align*}
with corresponding forms for $B$ and $D$. Neglecting terms quadratic in 
disturbance amplitude leaves a coupled system in $A$ and $C$, and a coupled system in $B$ and $D$. 
The latter system reduces to the former with the equivalence $A  
\leftrightarrow B, C \leftrightarrow D, \xibar \leftrightarrow \etabar, \omegadss_{bd} \leftrightarrow -\omegadss_{ac}$, so it is sufficient to just consider the behaviour of the coupled $A,C$ system, say, which reduces to solving  
 \begin{equation*}
    \begin{split}
     \left( \omegadss_{ac} + \kdl \vgdev - \frac{1}{2} \omega''_1 \kdl^2 + \mu_1
     |A_0|^2 - \frac{\mu_2 \amps^*}{|A_0|^2} \right) a_1 + \mu_1 |A_0|^2 a_2 + \mu_2 \left( |C_0|^2 + \frac{\amps^*}{|A_0|^2} \right) c_1 + \mu_2 |C_0|^2 c_2 & = 0, \\
     - \mu_1 |A_0|^2 a_1 + \left( \omegadss_{ac} + \kdl \vgdev +
\frac{1}{2} \omega''_1 \kdl^2 - \mu_1 |A_0|^2 + \frac{\mu_2 \amps}{|A_0|^2} \right) a_2 -
\mu_2 |C_0|^2 c_1 -\mu_2 \left( |C_0|^2 + \frac{\amps}{|A_0|^2} \right) c_2 & = 0, \\
     \mu_4 \left( |A_0|^2 + \frac{\amps}{|C_0|^2} \right) a_1 + \mu_4 |A_0|^2 a_2 + \left( \omegadss_{ac} 
- \kdl \vgdev -
     \frac{1}{2} \omega''_2 \kdl^2 + \mu_3 |C_0|^2 - \frac{\mu_4 \amps}{|C_0|^2} \right) c_1
     + \mu_3 |C_0|^2 c_2 & = 0, \\
     -\mu_4 |A_0|^2 a_1 - \mu_4 \left( |A_0|^2 + \frac{\amps^*}{|C_0|^2} \right) a_2 - \mu_3 |C_0|^2 c_1 + \left( \omegadss_{ac} - \kdl \vgdev + \frac{1}{2}
\omega''_2 \kdl^2 - \mu_3 |C_0|^2 + \frac{\mu_4 \amps^*}{|C_0|^2} \right) c_2 & = 0.
    \end{split}
   \end{equation*}
Thus, as in Section~\ref{s:cnls}, the dispersion relation may be obtained by setting a $4 \times 4$ determinant to be zero. However, in contrast to Section~\ref{s:cnls}, the presence of the terms involving $\amps$ makes it hard to derive analytically a quartic equation for $\omegadss_{ac}$. Thus,  
we find the roots numerically for a  given parameter set, by calling an eigenvalue routine for the above system, giving $\omegadss_{ac}$ directly. 


\section{Numerical simulations} \label{s:numerics}

We perform numerical simulations of the original system (\ref{1.2}) of
coupled {\sg} equations to examine the accuracy and
validity of the theoretical predictions of instability derived in
Sections~\ref{s:2wave} and~\ref{s:4wave}. 
In our previous study~\cite{Griffiths} we verified that the modulational instability of two-wave solutions is well described by the {\nlcnls} equations, over a wide parameter range. Here, we make a more detailed comparison between the predictions of this theory, those of the {\cnls} equations, and the numerical results, for the instability of both two-wave and four-wave solutions.  
Although our emphasis will be on verifying the growth rate of modulational instabilities, we will also comment on the structure of the solutions when any instabilities reach maximum amplitude.

\subsection{The numerical method}

Since the plane wave solutions and the modulational instabilities we analyse are spatially periodic, it is natural to use a Fourier basis for the $x$-dependence of $u$ and $w$.
Then, denoting a quantity $q(x,t)$ as $\sum_{j \ge 0} \tilde q_j(t) e^{ijx}~+~{\rm c.c.}$, where $j$ is one of the numerically resolved wavenumbers, we write (\ref{1.2}) as the first order system
    \begin{equation}
     \begin{pmatrix} \tilde u \\ {\tilde u}_t \\ \tilde w \\ \tilde 
w_t \end{pmatrix}_t =
     \begin{pmatrix}
            0 & 1 & 0 & 0 \\
             - \left( \delta^2 + j^2 \right) & 0 & \delta^2 & 0 \\
            0 & 0 & 0 & 1 \\
            1 & 0 & - \left( 1 + c^2 j^2 \right) & 0
       \end{pmatrix}
      \begin{pmatrix} \tilde u \\ \tilde u_t \\ \tilde w \\ \tilde w_t 
      \end{pmatrix}
       + \tilde N_j(t)
       \begin{pmatrix} 0 \\ \delta^2 \\ 0 \\ -1 \end{pmatrix},
     \elabel{scheme1}
    \end{equation}
where the terms  $\tilde N_j(t)$ originate from $N(x,t)=(u-w) - \sin(u-w)$. We solve this using a split-step 
scheme, a method which has frequently been applied to such nonlinear wave 
equations.
Neglecting the nonlinear terms in (\ref{scheme1}), \ie those 
involving $\tilde N$, we obtain a linear system that can be updated 
exactly, since given $\tilde u$, $\tilde u_t$, $\tilde w$ and $\tilde 
w_t$ at time $t$, and $\omega_{1,2}(j)$, we can write down explicitly 
the solution of the linear system at time $t+\Delta t$. Similarly, 
neglecting the linear terms in (\ref{scheme1}), \ie those not 
involving $\tilde N$, we obtain a system in which $\tilde u$ and 
$\tilde w$ are constant, so that $\tilde u_t$ and $\tilde w_t$ can be 
trivially updated by using the value of $\tilde N$ at time $t$. This 
latter property relies on the second derivatives with respect 
to $t$ in the governing equations.  By combining these two steps in a 
sequence of appropriately chosen fractional steps, one may obtain a 
method of any order (as explained, for instance, in section 2 
of~\cite{Muslu}). For the simulations to be presented, we use a fourth order scheme with a time-step in the range 0.025 to 0.1, with 96 to 768 points in the $x$-direction, and the term $\tilde N$ evaluated pseudospectrally. The nonlinear scheme then conserves energy to a high degree -- typically to within $10^{-6}\,\%$. 
If the nonlinear terms are neglected, then the scheme exactly resolves linear dynamics, 
within roundoff error.

For initial conditions, we take either (\ref{ics2}) or (\ref{ics4}) which lead to the two-wave or four-wave 
exchange solutions (\ref{ACvals}) and (\ref{ABCDvals}) respectively, with wavenumber $k$ and amplitude parameter $\amp$. The domain width $\lx$ is then restricted by the need to choose $\lx = 2 \pi n/k$, where $n$ is a positive integer. To trigger any instabilities, we add background noise to $u$ and $w$, at a level at least 1000 times smaller than the amplitude parameter $\amp$. 
In unstable cases, we identify a modulational instability of wavenumber $\kd$ by calculating the energy in pairs of sideband modes with wavenumbers $k \pm \kd$. To do this we consider the quantity
 \begin{equation}
   \eside{m} (t) = \tilde E_{k+\kd} + \tilde E_{k-\kd}, \; \; \kd = 2 \pi m / \lx, \; \; 
                              \mbox{with $m$ a positive integer,} 
   \elabel{eside} 
 \end{equation} 
corresponding to the energy in the $m$-th pair of numerically resolved sideband modes. 
Here $\tilde E_j$ is the energy, calculated according to (\ref{e}), when $u$ and $w$ are replaced by their Fourier components with wavenumber $j$. Initially, $\eside{m}(t)$ typically grows exponentially for one or more $m$, permitting growth rates to be calculated for the corresponding wavenumbers $\kd = 2 \pi m /\lx$. 
 
To check the relevance of these growth rates, we also distinguish a disturbance energy. Although the linear solution projects solely onto wavenumber $k$, the unperturbed nonlinear plane wave solution under consideration projects onto wavenumbers $(2n+1) k$, for all non-negative integers $n$. 
Therefore we define an energy associated with the plane wave and its nonlinear harmonics
 \begin{equation}
   \eplane = \sum_{n \ge 0} \tilde E_{(2n+1)k}, 
  \elabel{eharms}
 \end{equation}
and an energy associated with the disturbance field
 \begin{equation}
   \ed = \sum_{j \neq (2n+1) k} \tilde E_j. 
   \elabel{edist} 
 \end{equation}
An overall disturbance growth rate can then be calculated by analysing $\ed$. Typically, the growth rate of one pair of sideband modes approximately equals the growth rate of $\ed$, so that we can identify with some certainty which sideband pair yields the most relevant modulational instability. 

This method has been tested by calculating the growth rate for the 
modulational instability of a one-wave solution at $c=1$ and $\eps 
\ll 1$, when stability is determined by that of a corresponding 
uncoupled {\sg} equation. Calculated growth rates typically agree with
the well-known theoretical prediction (\eg \cite{Fauve}) to at least
four significant figures.

\subsection{Instability at small amplitude}

We first consider the instability of small amplitude plane-wave solutions, for which the individual wave amplitudes $\eps A_0$, $\eps B_0$, $\eps C_0$ and $\eps D_0$ are all less than 0.1. We reduce the parameter regime to manageable proportions by only considering waves that satisfy the full-exchange condition (\ref{2.3a}). Then, according to (\ref{ACvals}) and (\ref{ABCDvals}), and recalling from Section~\ref{s:exchange} that $\alpha_1=-\alpha_2=\delta^{-1}$ for full exchange parameters, we have $\eps A_0 = \eps C_0 = \amp/4$ for two-wave solutions, and $\eps A_0 = \eps B_0 = \eps C_0 = \eps D_0 = \amp/8$ for four-wave solutions. Here $\amp$ is the maximum amplitude of $u$ in the linear solution, although the maximum amplitude of $w$ is $\amp/\delta$. 

The results are summarised in Table~\ref{t:small}. 
Numerically determined growth rates $\sn$ are given for modulational instability at just a single wavenumber $\kn$, corresponding to a relatively strong instability in each case. 
For the two-wave runs $\amp$ has been chosen so that the maximum value of $u$ or $w$ in the linear solution is 0.2, so that the values of $\eps A_0$ and $\eps C_0$ are 0.025 or 0.05. For the four-wave runs, $\amp$ is twice that of the corresponding two-wave simulation, so that the maximum value of $u$ or $w$ in the linear solutions is now 0.4. However, the individual wave amplitudes are the same in both cases, and in this sense the runs are closely related. 

\begin{table}[h]
\begin{center}
\begin{tabular}{|c|c||c|c|c|c|c|c|}
\hline
 & $c$             
 & 1.5     & 1.1     & 0.5     & $\sqrt{3}$ &$\sqrt{2/3}$&$\sqrt{1.5}$\\
 & $\delta$        
 & 1.5     & 1.1     & 0.5     & 3          & 0.5          & 3 \\
Wave & $k$             
 & 1       & 1       & 1       & 2          & 1.5          & 4 \\
Parameters & $\vg{1}$        
 & 1.228   & 1.049   & 0.722   & 1.265      & 0.884        & 1.066 \\
& $\vg{2}$        
 & 0.746   & 0.607   & 0.472   & 1          & 0.722        & 0.945 \\
& $\vg{1}-\vg{2}$ 
 & 0.483   & 0.441   & 0.249   & 0.265      & 0.162        & 0.121 \\
\hline
 & $\amp$ 
 & 0.2     & 0.2     & 0.1     & 0.2        & 0.1          & 0.2 \\
 2-wave & $\kn$           
 & 1/12    & 0.1     & 0.05    & 1/15       & 0.075        & 0.125 \\
 results & $\sn$           
 & 0.00248 & 0.00268 & 0.00116 & 0.00224    & 0.00091      & 0.00168 \\
 & $\hat s_c$      
 & 0.00249 & 0.00268 & 0.00116 & 0.00221    & 0.00091      & 0.00167 \\
 & $\hat s_{ac}$   
 & 0.00249 & 0.00268 & 0.00116 & 0.00223    & 0.00091      & 0.00168 \\
\hline
 & $\amp$          
 & 0.4     & 0.4     & 0.2     & 0.4        & 0.2          & 0.4 \\
 4-wave & $\kn$           
 & 1/12    & 0.1     & 0.05    & 1/15       & 0.075        & 0.125 \\
 results & $\sn$           
 & 0.00243 & 0.00263 & 0.00116 & 0.00196    & 0.00089      & 0.00149 \\
 & $\hat s_c=\hat s_d$ 
 & 0.00248 & 0.00268 & 0.00120 & 0.00202    & 0.00089      & 0.00154 \\
 & $\hat s_{ac}=\hat s_{bd}$ 
 & 0.00248 & 0.00268 & 0.00119 & 0.00200    & 0.00089      & 0.00152 \\
\hline
\end{tabular}
\end{center}
\vspace*{-5mm}
\caption{Numerically determined growth rates $\sn$ for small amplitude solutions.}
\label{t:small}
\end{table}

We first discuss the two-wave results. There are two theoretical predictions for the growth rate. The first comes from the {\nlcnls} equations, valid when $\kd =O(\eps)$ and $(\vg{1}-\vg{2}) = O(1)$. This theory predicts the existence of two possible modes of instability, one evident as growth in wave amplitude $A$ and the other as growth in $C$, and can be diagnosed from (\ref{rates2}). Since $\omega''_1<0$ for all the parameters considered here, $\omegad_a$ is real and hence $A$ is always stable. Thus, we need only calculate the growth rate of $C$, which we denote by $\sd_c=\Im(\omegad_c)$. The second prediction comes from the coupled system (\ref{sw1}), valid when $\kd = O(\eps)$ and $(\vg{1}-\vg{2})=O(\eps)$, or equivalently from the {\cnls} equations. This predicts growth of a coupled mode in $A$ and $C$, and we denote this growth rate by $\sd_{ac}=\Im(\omegad_{ac})$.  

As shown in Table~\ref{t:small}, both $\sd_c$ and $\sd_{ac}$ are in good agreement with $\sn$ for all the parameters considered. Indeed, it is perhaps surprising that $\sd_c$ is a good approximation for the final one or two cases, when $\vg{1}-\vg{2}$ is apparently small. However, one can also see that $\sd_c$ and $\sd_{ac}$ are almost equal for these cases. 
This seems to be because for a wide range of full exchange parameters, such as those in Table~\ref{t:small}, at least one of $\mu_2$, $\mu_4$, $\omega''_1$ and $\omega''_2$ is small. Thus, for small amplitude conditions, the coupling term $R$ in (\ref{sw1}) is not important, so that even when $\vg{1}-\vg{2}$ is small the growth rate is approximately determined by the uncoupled instability in $C$.

Turning to the four-wave results, we once again have two predictions for the growth rate. 
The first is the uncoupled growth rate for the $C$ and $D$ modes, which have the same growth rate $\sd_c$. Even though $\sd_a=\sd_b$ can be positive, $\sd_c > \sd_a$ for all the parameters in Table~\ref{t:small}. The second prediction, denoted by $\sd_{ac}$, is no longer derived from the {\cnls} equations, but rather from the {\nlcnls} equations, valid when $\kd=O(\eps)$, with $(\vg{1}-\vg{2})=O(\eps)$, and $\vg{1},\vg{2}$ both $O(1)$, as described in Section~\ref{s:nlcnls4}. It now describes a coupled instability in $A$ and $C$, or an equivalent instability in $B$ and $D$, with the same growth rate.  

As shown in Table~\ref{t:small}, there is now generally a small discrepancy between the numerically determined growth rate and the best theoretical prediction. We speculate that this is an effect of the wave amplitude -- recall that in the linear solution the maximum amplitude reached by $u$ or $w$ is 0.4, twice that of the two-wave simulations. This point will be clarified in the next section.

\subsection{Instability at moderate amplitude} \label{s:modamp}

\begin{figure}[h]
\begin{center}
\begin{picture}(155,100)(0,0)
\put(-5,-5){\includegraphics[width=160mm]{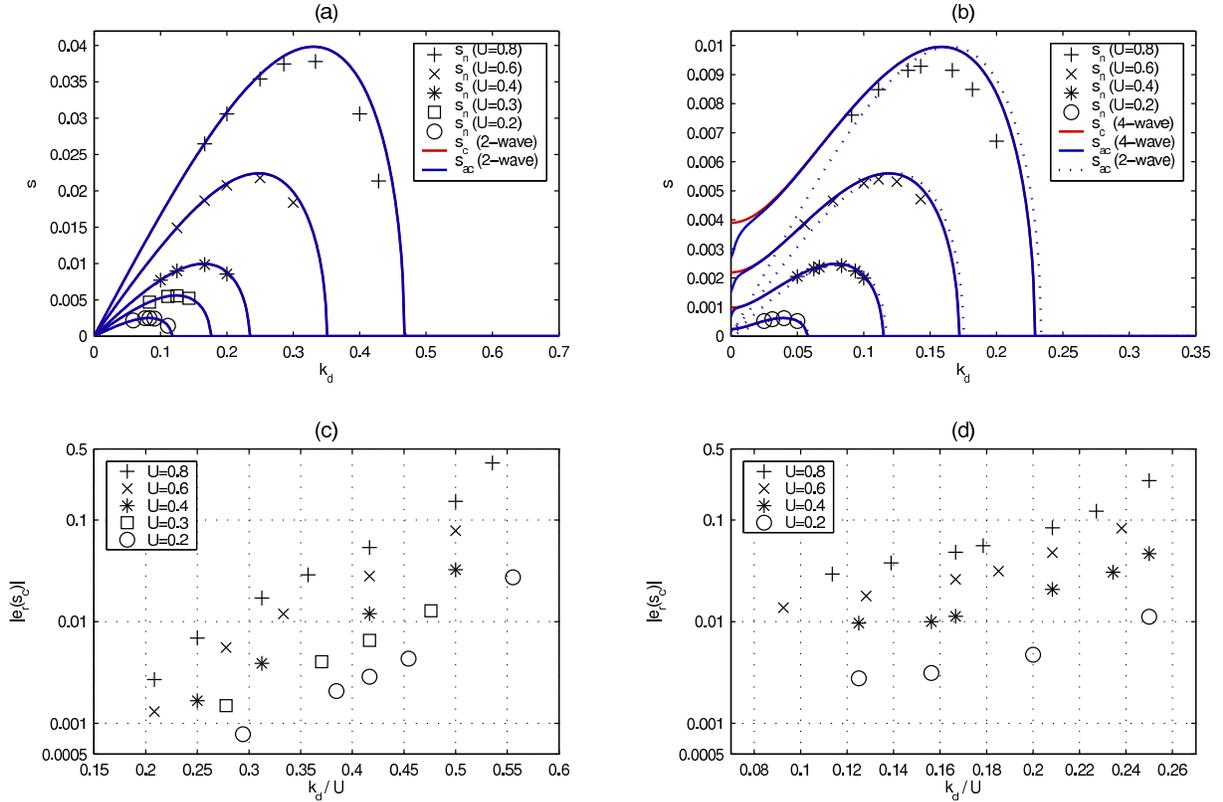}}
\end{picture}
\end{center}
\caption{{\small Numerically determined growth rates and theoretical predictions at $c=\delta=1.5$, $k=1$ as a function of disturbance wavenumber $k_d=\kd$ . (a) Two-wave results. (b) Four-wave results. (c) The quantity $|\er(\sd_c)|$ for the two-wave prediction $\sd_c$, as defined in (\ref{er}). (d) The quantity $|\er(\sd_c)|$ for the four-wave prediction $\sd_c$.}}
\label{f:par10}
\end{figure}

We now consider how $\sn$ and the functions $\sd_c$, $\sd_{ac}$ change as $\amp$ is increased, for three different parameter sets. 
We first discuss a full-exchange case with $c=\delta=1.5$, $k=1$, for which
$\vg{1}=1.23$, $\vg{2}=0.75$, so that $\vg{1}-\vg{2}=0.48$. 
The nonlinear interaction coefficients are $\mu_1=0.003$, $\mu_2=0.13$, $\mu_3=1.00$, $\mu_4=0.08$, and $\omega''_1=-0.31$, $\omega''_2=0.73$. 
The results are shown in Figure~\ref{f:par10}. For the two-wave case, given in Figure~\ref{f:par10}a, the theoretical predictions $\sd_c$ and $\sd_{ac}$ are almost indistinguishable. For small $\amp$ they closely follow $\sn$, whilst as $\amp$ is increased the differences between the predictions and $\sn$ grow. The four-wave results, given in Figure~\ref{f:par10}b, show the same type of behaviour. However, note how the four-wave predictions are different from the corresponding two-wave predictions (at $\amp/2$, as shown by the dotted line), particularly at small $\kd$. 

Although it is apparent from Figure~\ref{f:par10} that both $|\sd_c-\sn|$ and $|\sd_{ac}-\sn|$ vanish as $\amp \rightarrow 0$, we would really like to show that the relative error of the predictions vanishes as $\amp \rightarrow 0$, and hence that $\sd_c$ and $\sd_{ac}$ have the correct limiting behaviour at small amplitude. Therefore, given a numerically determined growth rate $\sn$, for a theoretical prediction $\sd_p$ we introduce a relative error
 \begin{equation}
   \er(\sd_p)=\frac{\sn-\sd_p}{\sn}. \elabel{er}
 \end{equation}
The quantity $|\er(\sd_c)|$ is shown in Figures~\ref{f:par10}c,d for the runs of  Figures~\ref{f:par10}a,b. In both the two-wave and four-wave cases one can see that the relative error does vanish as $\amp \rightarrow 0$. Indeed, one can show in each case that $\er \approx \amp^2 \, \times$ some function of $(\kd/\amp)$. Note that the relative errors for the two-wave and four-wave cases are of the same order of magnitude when comparing by $\amp$ (rather than by the individual wave amplitude $\eps C_0$). 

For the two-wave cases we have already noted that $\sd_c$ and $\sd_{ac}$ are almost indistinguishable. However, is $\sd_{ac}$ more accurate than $\sd_c$? Perhaps, for the larger amplitude cases in Figure~\ref{f:par10}a, (\ref{ABCDeq4}) is a more suitable model than (\ref{ABCDeq3}) because the wave amplitudes are now of the same order as $(\vg{1}-\vg{2})=0.48$? Or perhaps the \cnls\ model (\ref{ABCDeq5}) is necessarily more accurate than the \nlcnls\ model (\ref{ABCDeq3})? At least for these parameters, the answer to all three of these questions is no, since $|\sn-\sd_c|$ is marginally smaller than $|\sn-\sd_{ac}|$. Further, from (\ref{betasq}) we calculate that $|\vg{2}-\vg{1}| > \beta$ for all the two-wave cases, so that at the wavenumber $\kcc$ where the uncoupled C-mode becomes unstable the \cnls\ equations predict the destabilizing scenario 1(a) of Section~\ref{s:cnls}. However, one can clearly see in the large amplitude cases of Figure~\ref{f:par10}a that there is a stabilization at $\kd=\kcc$, rather than a destabilization.

\begin{figure}[h]
\begin{center}
\begin{picture}(155,100)(0,0)
\put(-5,-5){\includegraphics[width=160mm]{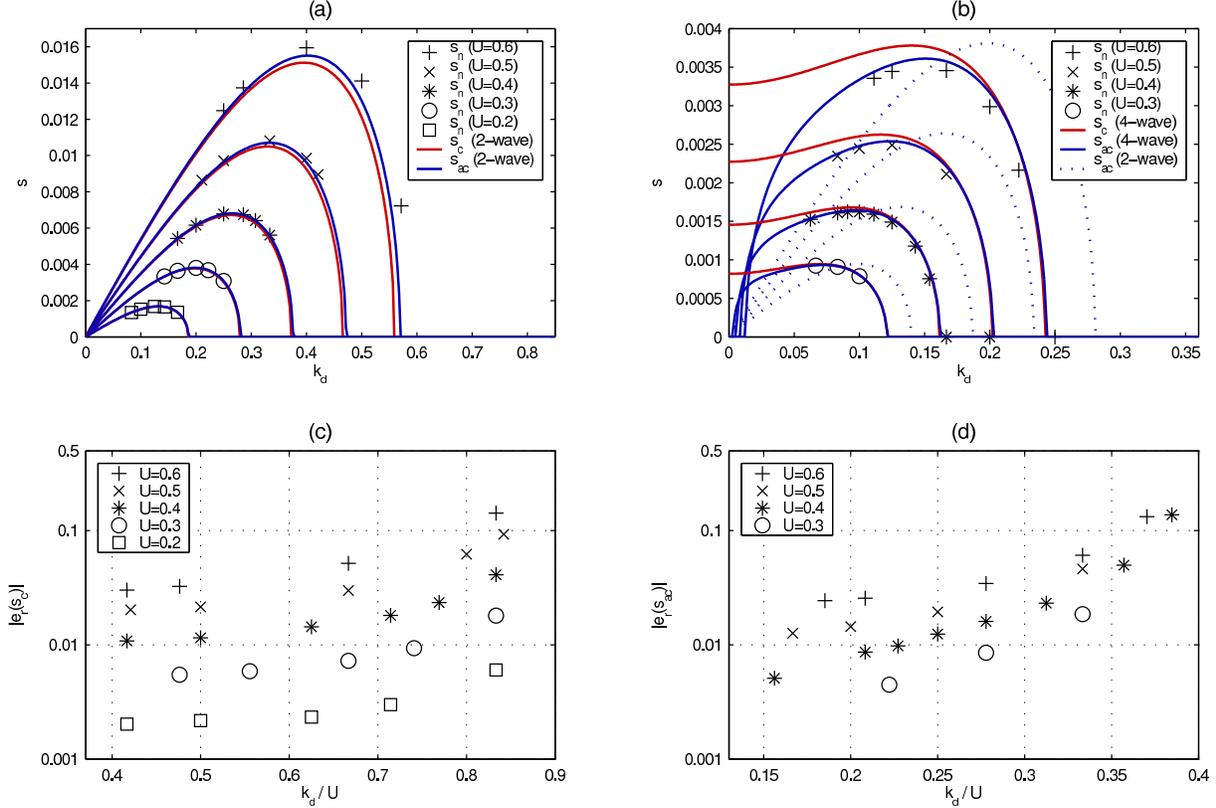}}
\end{picture}
\end{center}
\caption{{\small Numerically determined growth rates and theoretical predictions at $c=\sqrt{1.5}$, $\delta=3$, $k=4$ as a function of disturbance wavenumber $k_d=\kd$. (a) Two-wave results. (b) Four-wave results. (c) The quantity $|\er(\sd_{c})|$ for the two-wave prediction $\sd_{c}$, as defined in (\ref{er}). (d) The quantity $|\er(\sd_{ac})|$ for the four-wave prediction $\sd_{ac}$.}}
\label{f:par1}
\end{figure}

\begin{figure}[h]
\begin{center}
\begin{picture}(155,100)(0,0)
\put(-5,-5){\includegraphics[width=160mm]{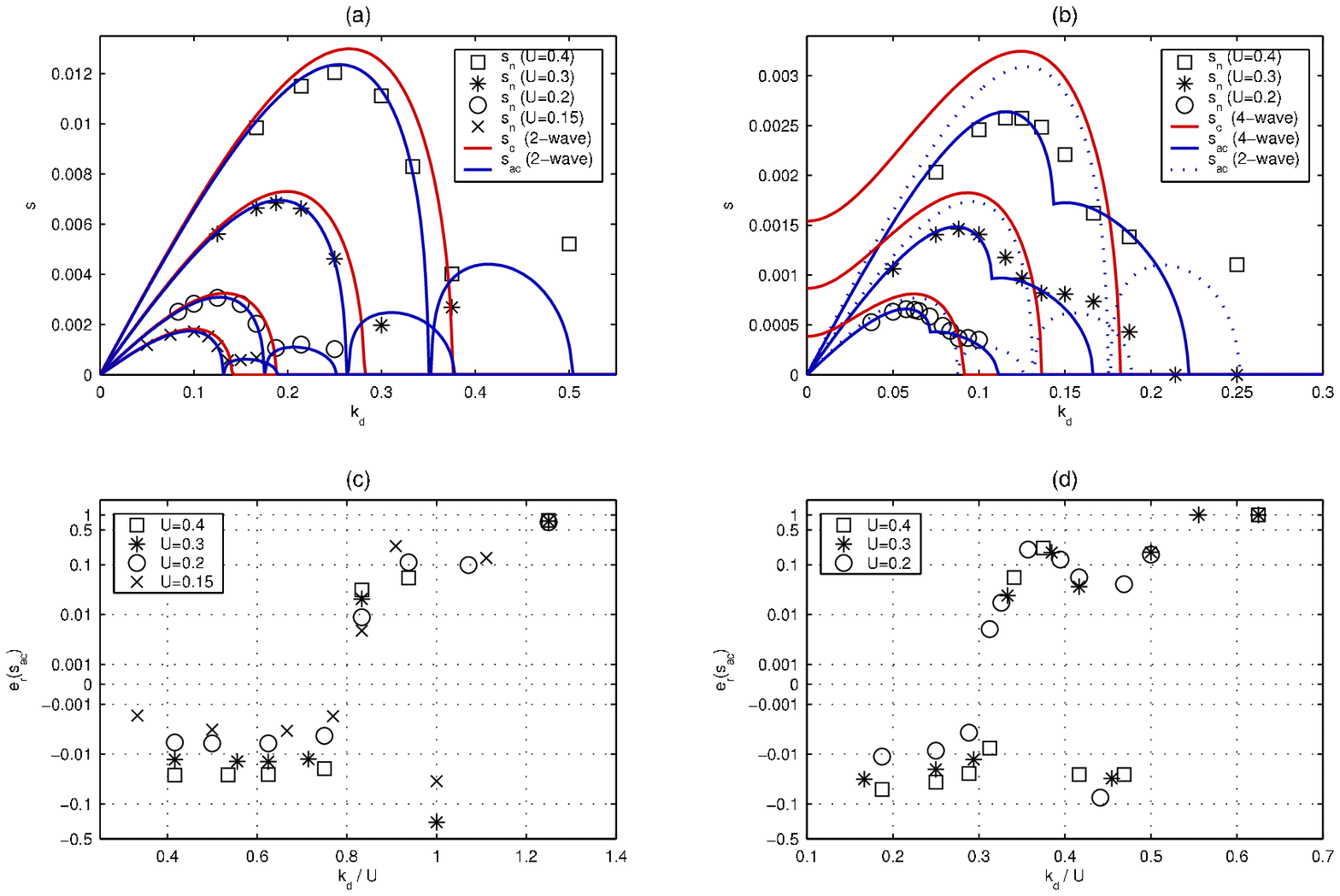}}
\end{picture}
\end{center}
\caption{{\small Numerically determined growth rates and theoretical predictions at $c=2/3$, $\delta=3/4$, $k=3/2$ as a function of disturbance wavenumber $k_d=\kd$ . (a) Two-wave results. (b) Four-wave results. (c) The quantity $\er(\sd_{ac})$ for the two-wave prediction $\sd_{ac}$, as defined in (\ref{er}). (d) The quantity $\er(\sd_{ac})$ for the four-wave prediction $\sd_{ac}$.}}
\label{f:par8}
\end{figure}

The second case we discuss, also a full exchange solution, has $c=\sqrt{1.5}$, $\delta=3$, $k=4$, for which
$\vg{1}=1.07$, $\vg{2}=0.95$, so that $\vg{1}-\vg{2}=0.12$. The nonlinear interaction coefficients are $\mu_1=0.05$, $\mu_2=0.38$, $\mu_3=0.67$, $\mu_4=0.34$, and $\omega''_1=-0.12$,
$\omega''_2=0.19$. The results are shown in Figure~\ref{f:par1}. For the two-wave case, both $\sd_c$ and $\sd_{ac}$ offer a good approximation to $\sn$. At small amplitudes, $\sd_c$ and $\sd_{ac}$ are almost identical, as expected, and as shown in Figure~\ref{f:par8}c, $\sd_c$ shows the desired convergence properties as $\amp \rightarrow 0$. At larger amplitudes, where perhaps we enter the regime with $\amp \sim (\vg{1}-\vg{2})$, there are differences between $\sd_c$ and $\sd_{ac}$, and here $\sd_{ac}$ is a better approximant. 
Indeed, $|\vg{2}-\vg{1}| > \beta$ as defined in (\ref{betasq}), so that the \cnls\ models predict the destabilizing scenario 1(a) of Section~\ref{s:cnls} with $\sn>0$ at $\kd=\kcc$, which is verified at $\amp=0.6$, for instance.
For the four-wave case, there are clear differences between $\sd_c$ and $\sd_{ac}$ even at small amplitudes when $\kd$ is also small. To obtain relative errors comparable to the two-wave case, it is now necessary to use the prediction $\sd_{ac}$, as shown in Figure~\ref{f:par8}d. In both two-wave and four-wave cases, note that $|\er|$ increases with $\kd$ at fixed $\amp$, which is perhaps associated with a breakdown in the scale assumption implicit in the weakly nonlinear theory leading to $\sd_c$ and $\sd_{ac}$.  

Differences between the theoretical predictions $\sd_c$ and $\sd_{ac}$ can be highlighted by considering parameters such that $\vg{1}=\vg{2}$. Therefore we consider the case 
$c=2/3$, $\delta=3/4$, $k=3/2$, for which $\vg{1}=\vg{2}=0.71$. The nonlinear interaction coefficients are $\mu_1=0.07$, $\mu_2=0.29$, $\mu_3=0.60$, $\mu_4=0.57$, and $\omega''_1=-0.18$, $\omega''_2=0.37$. These do not satisfy the full exchange condition (\ref{2.3a}). For two-wave solutions, (\ref{ACvals}) gives $\eps A_0=0.13 \, \amp$, $\eps C_0=0.37 \, \amp$, and for four-wave solutions (\ref{ABCDvals}) gives $\eps A_0 = \eps B_0=0.065 \, \amp$, $\eps C_0 = \eps D_0 = 0.18 \, \amp$. According to (\ref{2.3}) and (\ref{2.6}), the maximum amplitude of $w$ in the linear solution is $1.17 \, \amp$. 

Results for these parameters are shown in Figure~\ref{f:par8}. For both two-wave and four-wave cases, there are now significant differences between $\sd_c$ and $\sd_{ac}$. Although $\sd_c$ gives  moderate agreement with $\sd_n$, it completely misses the second tongue of instability at large $\kd$, and $|\er(\sd_c)|$ does not decrease with $\amp$. However, $\sd_{ac}$ does have the correct qualitative behaviour. To assess its quantitative performance, we consider the relative error $\er(\sd_{ac})$, as shown in Figures~\ref{f:par8}c,d. Because $\sd_{ac}$ can be greater or less than $\sn$, we plot $\er(\sd_{ac})$ (which may be positive or negative) rather than $|\er(\sd_{ac})|$ as before, allowing the form of the curves to be fully appreciated.   
For the two-wave case, the points with $\kd/\amp \le 5/6$, which includes the region of maximum growth, show good convergence as $\amp \rightarrow 0$. For $\kd / \amp >  5/6$, convergence is no longer clear, although the actual behaviour could be masked by the limited discrete resolution of $\er$. Nevertheless, errors here remain at about $10\%$. For the four-wave case, the points with $\kd/\amp \le 0.32$, which includes the region of maximum growth, also show good convergence as $\amp \rightarrow 0$. For $\kd / \amp > 0.32$, it appears that $\er(s_{ac})$ assumes a form independent of $\amp$, with no decrease as $\amp \rightarrow 0$, although errors here still remain at about 10\%. 

For the two-wave cases we have $\beta > |\vg{2}-\vg{1}| = 0$ as defined in (\ref{betasq}), so that we lie in scenario 1(b) of Section~\ref{s:cnls}.
One can then verify that $R < R_c$ for all the cases in Figure~\ref{f:par8}a, so that the \cnls\ models predict a destabilization at $\kd=\kcc$, as is clearly the case at each $\amp$. Note this destabilization at $\kd=\kcc$ is not inconsistent with the vanishing of $\sd_{ac}$ (and $\sn$) for some $\kd < \kcc$, as noted at the end of scenario 1(b). This twin-tongued growth rate structure is typical for two-wave cases with $|\vg{1}-\vg{2}| \ll 1$. 
 
For the four-wave cases, the numerical simulations also show the existence of a second tongue of instability with a higher disturbance wavenumber than those shown in Figure~\ref{f:par8}. This appears at 
$\amp=0.3$ as an instability with $\sn=0.00256$ and $\kd=15/34$, and at $\amp=0.4$ as an instability with $\sn=0.00483$ and $\kd=3/7$, and in both cases is the strongest instability observed at that amplitude $\amp$. Such instabilities are not predicted by any of our theories. Indeed, even though this is an instability of a weakly nonlinear plane wave, the modulating wavenumber $\kd \approx k/3$, so that it does not correspond to a long spatial modulation of the type considered here.

\subsection{Character of the instability at finite amplitude}

In our previous study of the instability of two-wave solutions~\cite{Griffiths}, we showed how the energy exchange between $u$ and $w$ in the plane-wave solution can have a counterpart during the finite amplitude stage of a modulational instability. We now examine further the nature of this finite amplitude energy exchange, and extend the discussion to the four-wave case and to the long-time behaviour of the system. 

In order to make clear conclusions, we consider a restricted class of scenarios. First, we restrict the plane-wave parameters to $c=\delta$ and $k=1$, all of which satisfy the full exchange condition (\ref{2.3a}). Then, for $c$ and $\delta$ of $O(1)$, $\vg{1}-\vg{2} \approx 0.5$, and we lie in a stability regime like that of Figure~\ref{f:par10}. Second, we restrict the computational geometry so that just one of the numerically resolved sidebands is unstable. We achieve this by taking $\lx \approx 2 \pi / \kd_{c*}$, where $\kd_{c*}$ is the most unstable wavenumber associated with the prediction $\sd_c$, as determined by (\ref{rates2}) or by the C-mode version of (\ref{rates4a}). We then obtain an evolution in which the initial plane-wave structure becomes modulated on the lengthscale of the computational domain, with no possibility of an interaction between modulational instabilities of two different wavelengths.  

We start by discussing an important special case, that of $c=\delta=1$. Here, the system (\ref{1.2}) decouples into a \sg\ equation and a linear wave equation, and, therefore, is integrable~\cite{Khus}. As a consequence, the modulational instability of plane-wave solutions shows a recurrence, similar to the famous Fermi--Pasta--Ulam recurrence~\cite{FPU}. 
We view this behaviour by considering the time evolution of three quantities: $\eplane$, as defined in (\ref{eharms}), representing the strength of the plane-wave and its nonlinear harmonics; $\eside{1}$, as defined in (\ref{eside}), representing the strength of the modulational instability in the first sideband pair, with disturbance wavenumber $\kd=2\pi/\lx$; and $E-\eplane-\eside{1}$,  representing the contribution from other wavenumbers. 

\begin{figure}
\begin{center}
\begin{picture}(155,120)(0,0)
\put(-5,-5){\includegraphics[width=160mm]{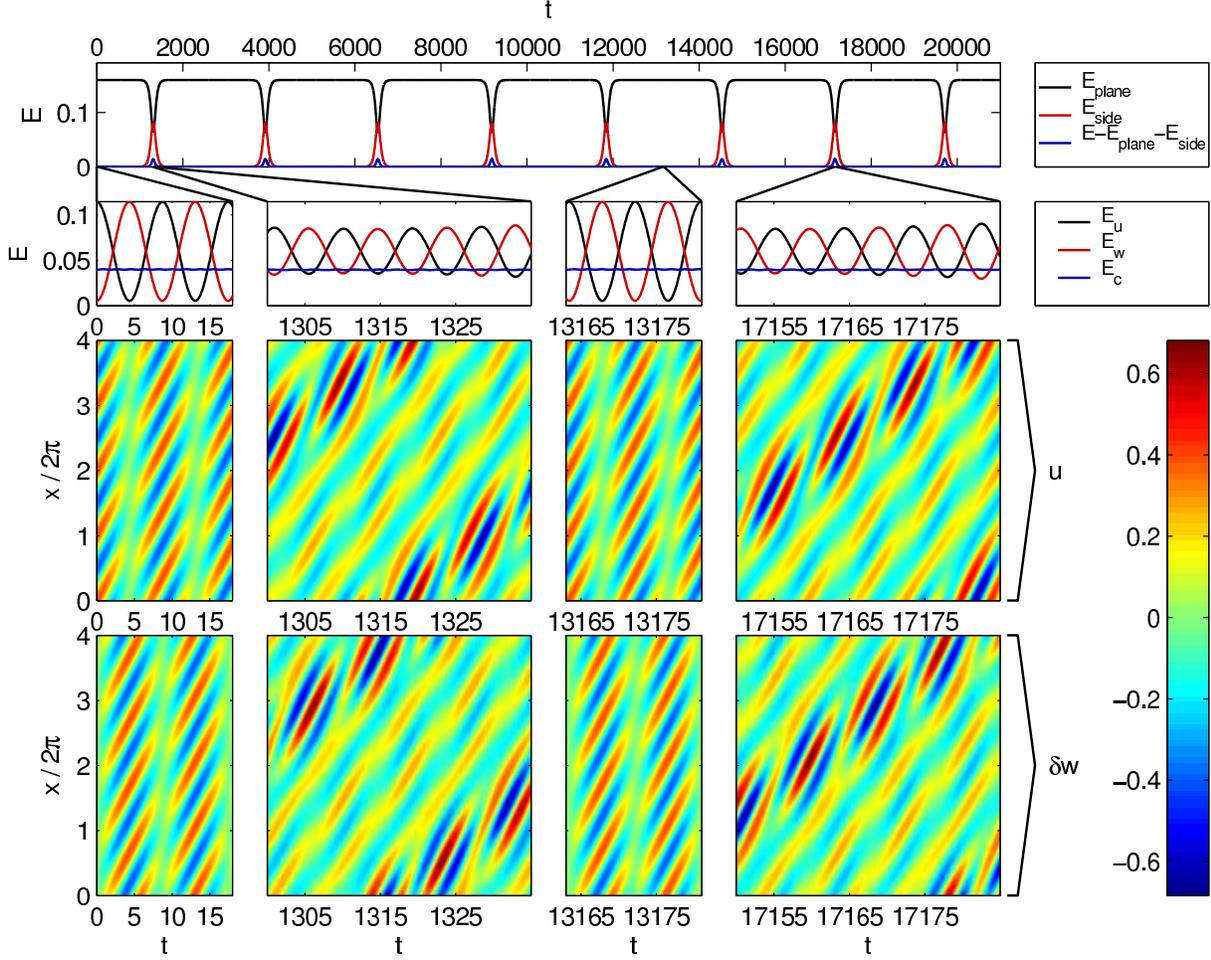}}
\end{picture}
\end{center}
\caption{{\small Time evolution of a two-wave solution at $c=\delta=1$, $k=1$, $\amp=0.4$, with $\lx=8\, \pi$.}}
\label{f:pars20-2}
\end{figure}

These quantities are shown in the top row of Figure~\ref{f:pars20-2}, during the instability of a two-wave solution at $c=\delta=1$, with $\amp=0.4$ and $\lx = 8 \pi$. One can clearly see an almost periodic recurrence, with most of the energy either in $\eplane$, or $\eside{1}$. In the second row, one can see how $E$ is partitioned between $u$, $w$ and the coupling, as defined in (\ref{e}), whilst in the third and four rows one can see $u(x,t)$ and $w(x,t)$ respectively (the complete computational domain is shown). At first the plane-wave solution has a periodic energy exchange, which is almost complete between $E_u$ and $E_w$, as  approximately described by (\ref{2.4}) and (\ref{2.5}). At the first minimum in $\eplane$, near $t=1300$, there is still a periodic energy exchange between $u$ and $w$, although the exchange is no longer almost complete, \ie $E_u$ and $E_w$ no longer almost vanish periodically. Further, at this time $u(x,t)$ and $w(x,t)$ are strongly modulated on the largest resolved scale, and the large amplitude behaviour pulses between $u$ and $w$. Thus, the almost complete energy exchange with spatially uniform structures initially observed is modified in the nonlinear regime to an incomplete energy exchange with structures localized on the lengthscale of the computational domain. Note that these same two scenarios alternate, in line with the recurrence of $\eplane$ and $\eside{1}$. One can see how $u(x,t)$ and $w(x,t)$ return almost to their initial plane-wave form when $\eside{1}$ is minimised, near $t=13,170$ for instance, whilst the pulsing modulations return when $\eside{1}$ is maximised, near $t=17,150$ for instance.      

\begin{figure}
\begin{center}
\begin{picture}(155,120)(0,0)
\put(-5,-5){\includegraphics[width=160mm]{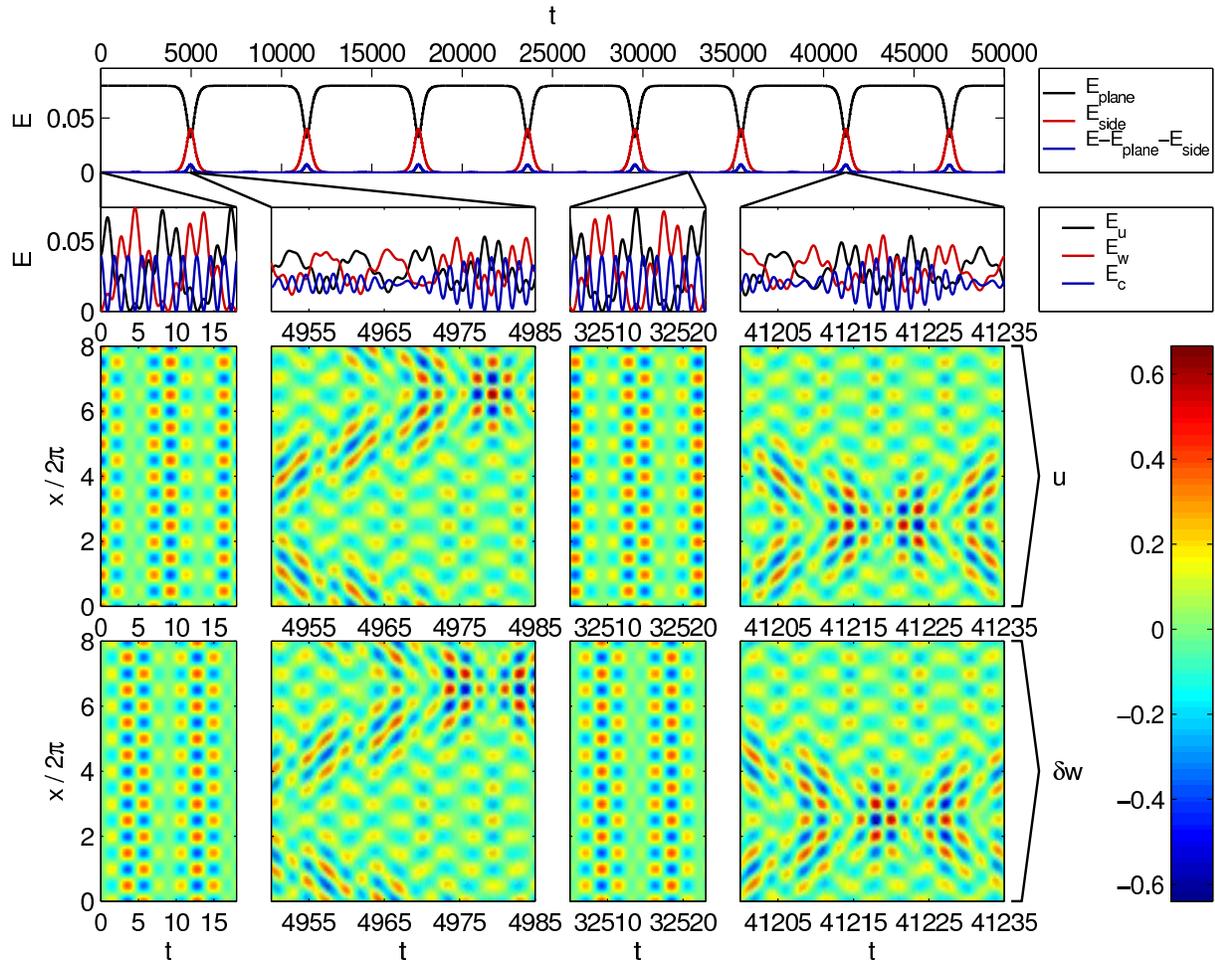}}
\end{picture}
\end{center}
\caption{{\small Time evolution of a four-wave solution at $c=\delta=1$, $k=1$, $\amp=0.4$, with $\lx=16\,\pi$.}}
\label{f:pars20-4}
\end{figure}

The four-wave case at $c=\delta=1$ with $\amp=0.4$ is similar, as shown in Figure~\ref{f:pars20-4}. Once again there is an almost periodic recurrence in $\eplane$ and $\eside{1}$. However, the partitioning of $E$ between the components $E_u$, $E_w$ and $E_c$ is more complicated than in the two-wave case, with rapid oscillations in each component superimposed on a slower exchange between $E_u$ and $E_w$. Once again, when $\eplane$ is maximised the slow exchange between $E_u$ and $E_w$ is almost complete, whilst when $\eside{1}$ is maximised the slow exchange becomes incomplete.   
The general pattern of an almost periodic sequence of alternating exchange patterns remains, from approximately uniform wavetrains to localized structures and back, and so on. 

\begin{figure}
\begin{center}
\begin{picture}(155,108)(0,0)
\put(-5,-5){\includegraphics[width=160mm]{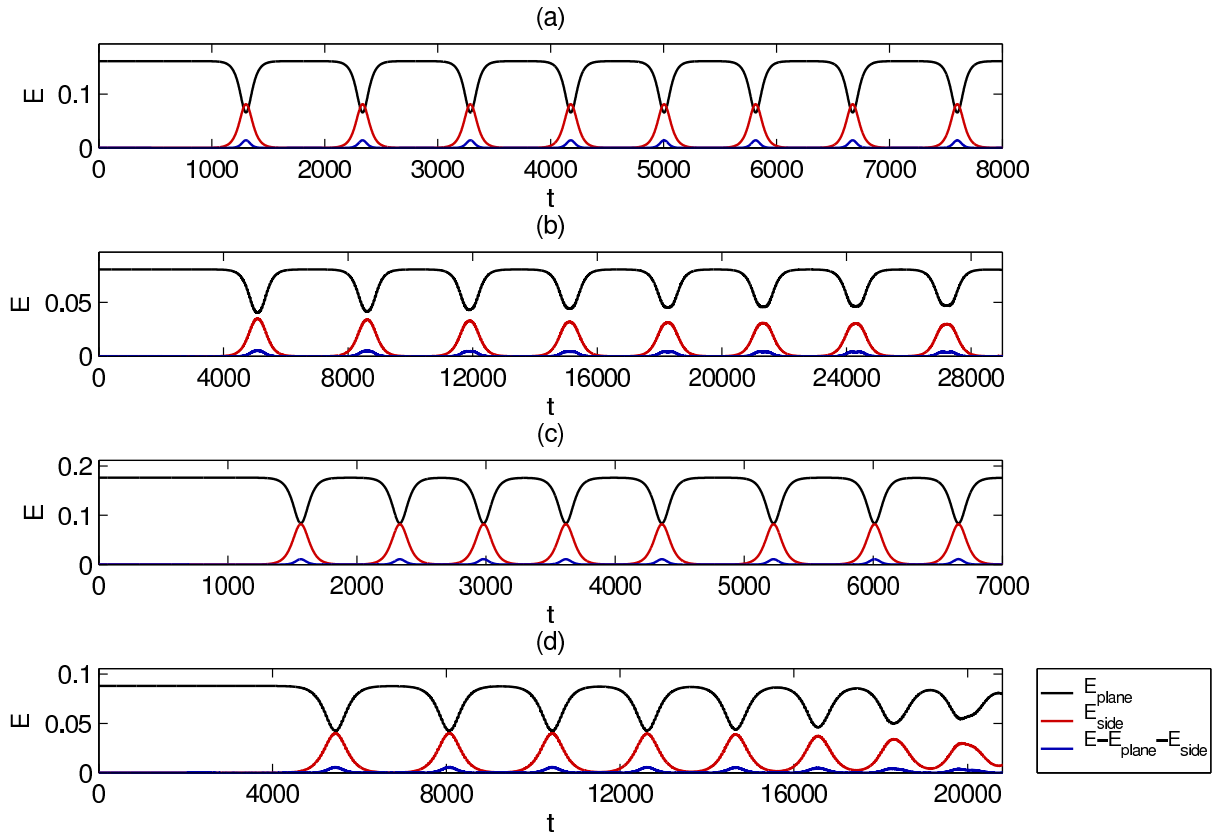}}
\end{picture}
\end{center}
\caption{{\small Time evolution of solutions with $k=1$ and $\amp=0.4$: (a) a two-wave solution at $c=\delta=1.01$, with $\lx=8\,\pi$; (b) a four-wave solution at $c=\delta=1.01$, with $\lx=16\pi$; (c) a two-wave solution at $c=\delta=1.1$ with $\lx=8\,\pi$; (d) a four-wave solution at $c=\delta=1.1$ with $\lx=16\,\pi$.}}
\label{f:cneq1}
\end{figure}

A natural question to ask is to what extent similar behaviour occurs when $c \neq 1$. 
Shown in Figure~\ref{f:cneq1} are results of corresponding simulations at $c=\delta=1.01$ and $c=\delta=1.1$, with $\amp=0.4$ in both cases. For the two-wave cases, shown in panels (a) and (c), the period of the recurrence now varies, although the character of each recurrence event appears to remain almost the same. For the four-wave cases, shown in panels (b) and (d), the character of the recurrence events now varies, but the evolution is still predominantly characterized by an exchange between $\eplane$ and $\eside{1}$. Thus both two-wave and four-wave cases exhibit an imperfect recurrence, which we might expect since, for $c$ close to 1, the equations are close to being an integrable system. Interestingly, the system (\ref{1.2}) with $c$ close to 1 has been used to model some dynamical processes in the DNA \cite{Yom,Yak}, where such energy exchange processes could, potentially, play an important role.

\begin{figure}
\begin{center}
\begin{picture}(155,120)(0,0)
\put(-5,-5){\includegraphics[width=160mm]{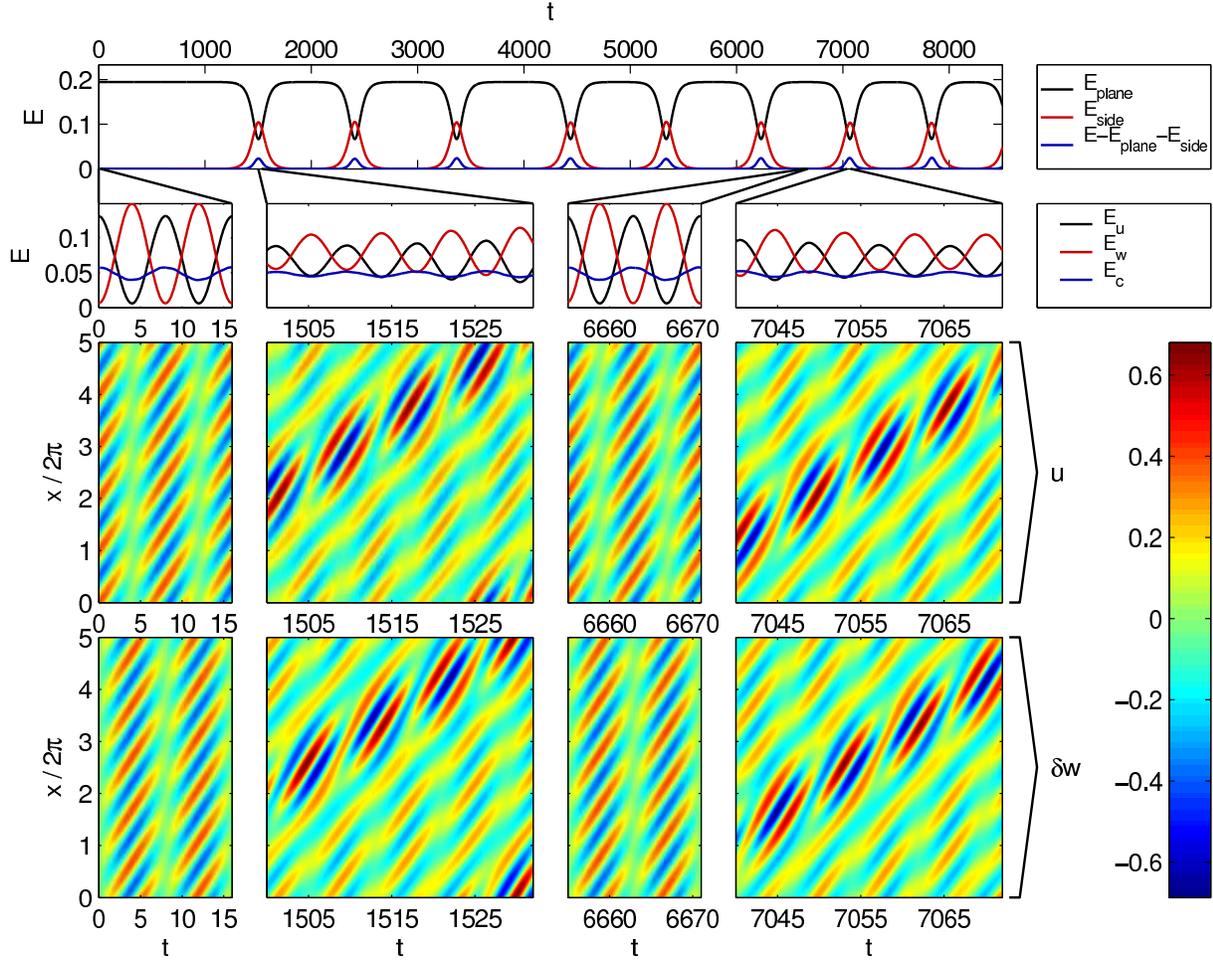}}
\end{picture}
\end{center}
\caption{{\small Time evolution of a two-wave solution at $c=\delta=1.2$, $k=1$ and $\amp=0.4$, with $\lx=10\,\pi$.}}
\label{f:pars12-2}
\end{figure}

\begin{figure}
\begin{center}
\begin{picture}(155,120)(0,0)
\put(-5,-5){\includegraphics[width=160mm]{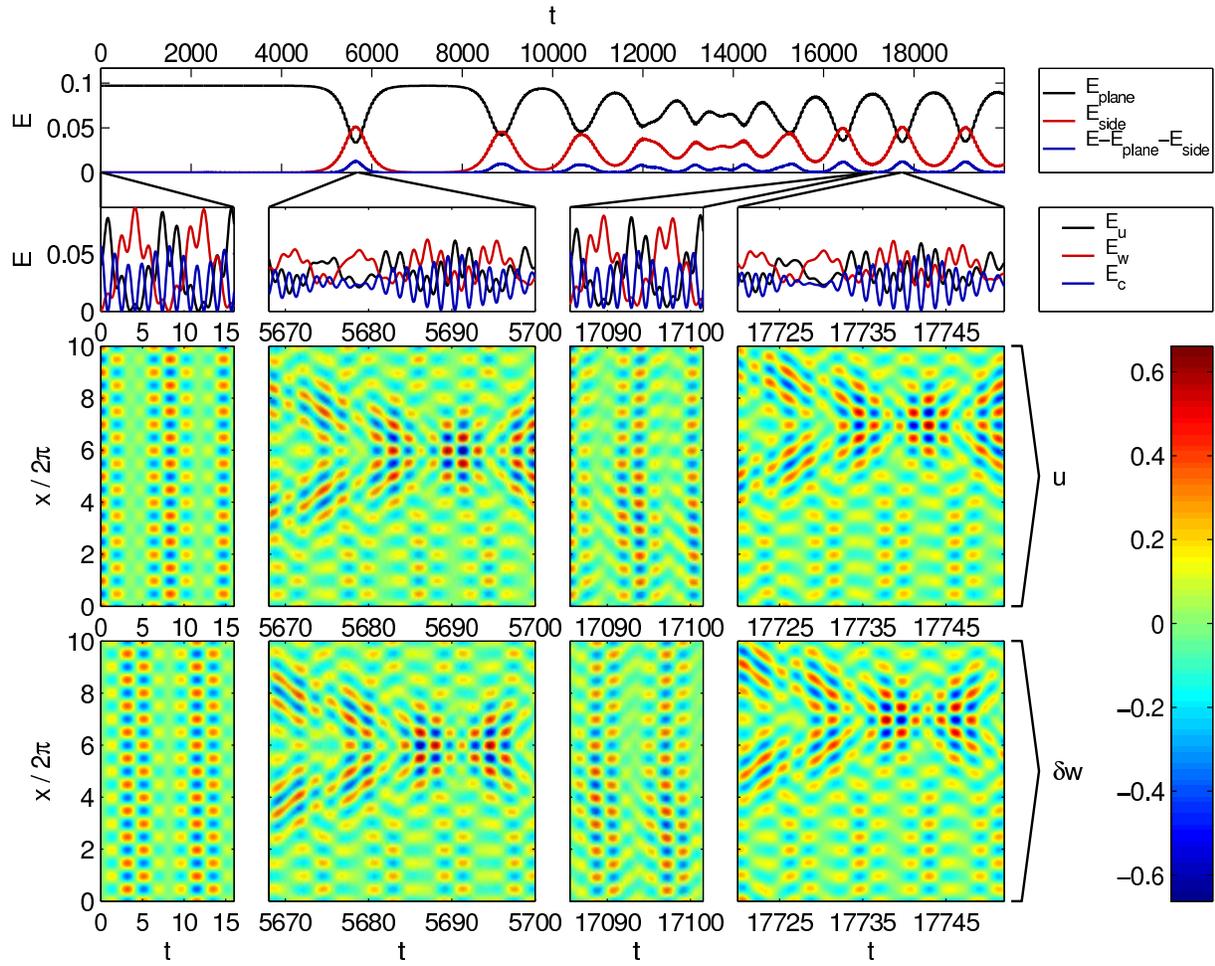}}
\end{picture}
\end{center}
\caption{{\small Time evolution of a four-wave solution at $c=\delta=1.2$, $k=1$ and $\amp=0.4$, with $\lx=20\,\pi$.}}
\label{f:pars12-4}
\end{figure}

Although the recurrence becomes imperfect when $c \neq 1$, the alternating pattern of  approximately uniform wavetrains and localized structures remains. Shown in Figures~\ref{f:pars12-2} and~\ref{f:pars12-4} are the results of  two-wave and four-wave simulations at $c=\delta=1.2$, with $\amp=0.4$. The two-wave case of Figure~\ref{f:pars12-2} still looks remarkably similar to that of Figure~\ref{f:pars20-2}, taken at $c=\delta=1$. However, in the four-wave case of Figure~\ref{f:pars12-4}, the recurrence breaks down around $t=13,000$, only to reappear later in the evolution. Further, during the second sequence of recurrence events, around $t=17,000$, the alternating pattern of approximately uniform wavetrains and localized structures reappears. For instance, around $t=17,100$ one can see that the solution only has a small spatial modulation, whilst around $t=17,730$ the solution has a spatial character almost identical to that of the first recurrence event around $t=5680$. More remarkably, note from the second row of Figure~\ref{f:pars12-4} how even the details of the exchange between $E_u$, $E_w$ and $E_c$ are almost identical around $t=5680$ and $t=17,730$.   
 
\begin{figure}
\begin{center}
\begin{picture}(155,120)(0,0)
\put(-5,-5){\includegraphics[width=160mm]{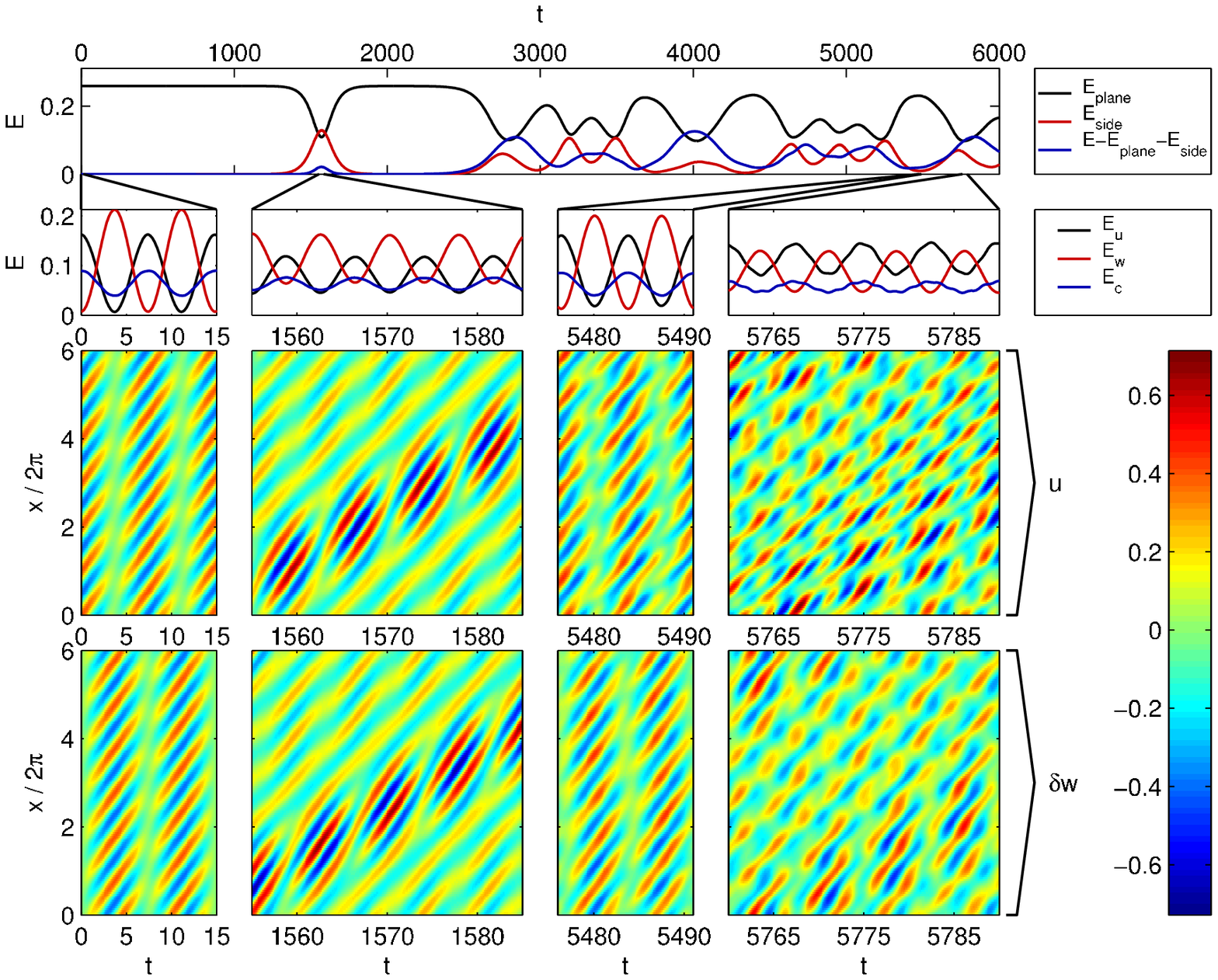}}
\end{picture}
\end{center}
\caption{{\small Time evolution of a two-wave solution at $c=\delta=1.5$, $k=1$ and $\amp=0.4$, with $\lx=12\,\pi$.}}
\label{f:pars10-2}
\end{figure}

\begin{figure}
\begin{center}
\begin{picture}(155,120)(0,0)
\put(-5,-5){\includegraphics[width=160mm]{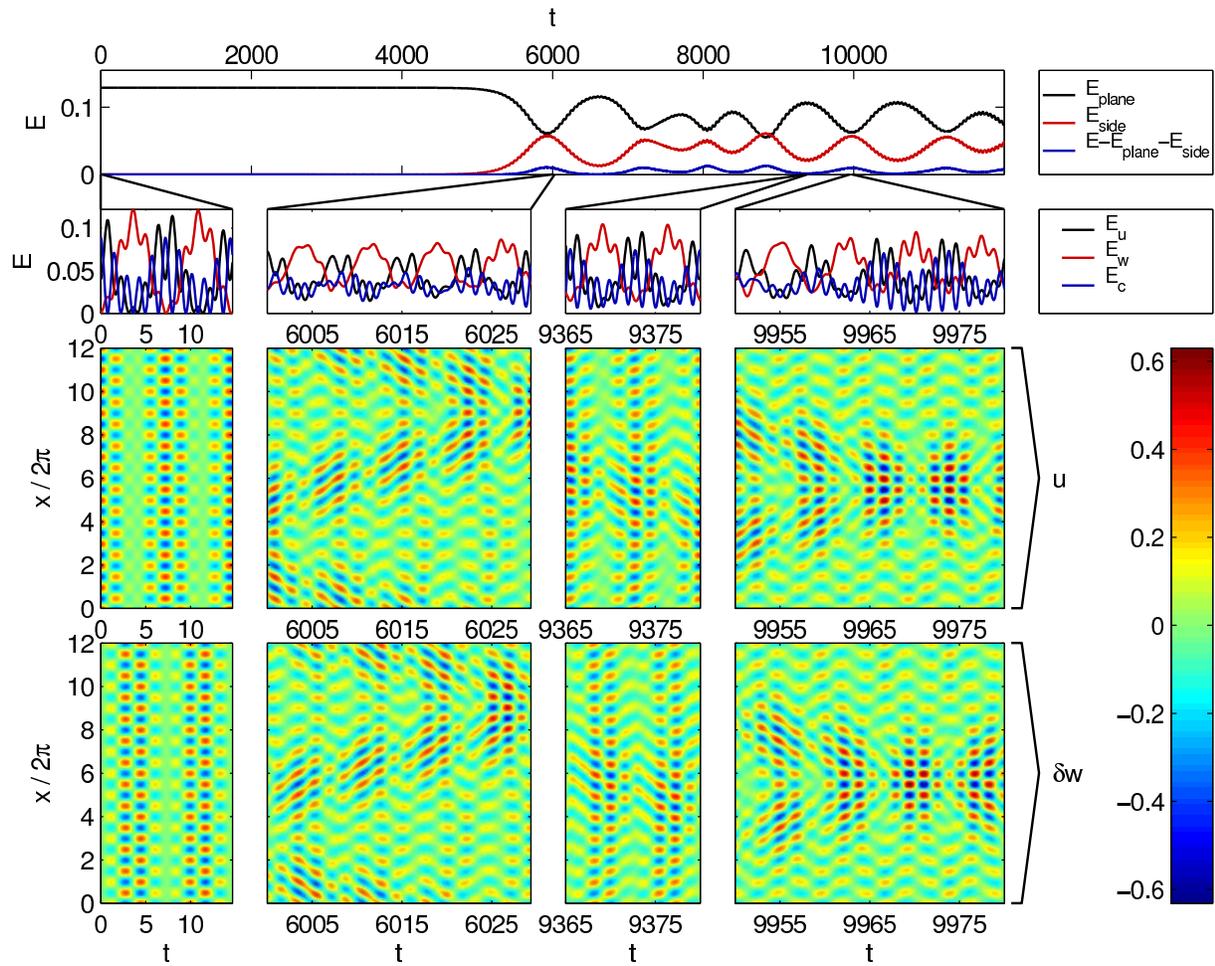}}
\end{picture}
\end{center}
\caption{{\small Time evolution of a four-wave solution at $c=\delta=1.5$, $k=1$ and $\amp=0.4$, with $\lx=24\,\pi$.}}
\label{f:pars10-4}
\end{figure}

However, for larger values of $c=\delta$, evolutions which might be described as recurrent become rare, and typically the solutions display a more irregular behaviour. To illustrate this, we show results for the case $c=\delta=1.5$.  Shown in Figure~\ref{f:pars10-2} is the evolution of a two-wave solution, with $\amp=0.4$ and $\lx= 12 \, \pi$, whilst shown in Figure~\ref{f:pars10-4} is the evolution a four-wave solution, with $\amp=0.4$ and $\lx=24 \, \pi$. 
(One can pick out these parameters in Figure~\ref{f:par10}, and verify that they correspond to relatively strong instabilities in each case).   
In the two-wave case, at large times the solution loses much of the character of the earlier recurrence.  
The four-wave case displays an irregular, imperfect recurrence, the character of which does seem to be similar to the earlier events. 

\section{Conclusion} \label{s:conclu}
 
We have studied the nonlinear dynamics of two pairs of counter-propagating waves within the framework of a two-component system of coupled \sg\ equations (\ref{1.2}). 
Using both analytical and numerical methods, we considered the evolution of spatially periodic solutions characterized by a single dominant wavenumber $k$, focussing on the possible modulational instability of spatially uniform plane-wave solutions. We concentrated on a particular class of plane-wave solutions corresponding to an energy exchange between the two physical components of the system, and for these solutions we noted how modulational instabilities can lead to a modification of the energy exchange, and to the formation of localized structures.    

The analytical results were based upon \ams\ expansions for the evolution of weakly nonlinear wave packets, which reduced (\ref{1.2}) to coupled equations for the evolution of four spatially varying wave amplitudes.  
However, the form of the amplitude equations depends crucially upon the assumed length-scale of the wave packets {\it vis-a-vis} the wave amplitude parameter $\eps \ll 1$ and the group velocities $\vg{1}$ and $\vg{2}$. In Section~\ref{s:models} we derived two classes of evolution equations for the regime relevant to this system with $\vg{1}$ and $\vg{2}$ both of $O(1)$: a set (\ref{ABCDeq1}) of {\nondis} equations for modulations on a lengthscale $\sim \eps^{-2}$, and two sets (\ref{ABCDeq3}) and (\ref{ABCDeq4}) of {\nlcnls} equations for modulations on a lengthscale $\sim \eps^{-1}$, according as to whether $(\vg{1}-\vg{2}) = O(1)$ or $O(\eps)$.
We have also considered a set (\ref{ABCDeq5}) of \cnls\ equations  with asymptotically weak dispersion, which formally includes each of the other models as different asymptotic reductions.
However, it is not clear which set of amplitude equations most efficiently captures the nonlinear dynamics of interest.  

The stability of plane-wave solutions involving only a single pair of waves was considered in Section~\ref{s:2wave}, whilst the stability of plane-wave solutions involving two pairs of counter-propagating waves was considered in Section~\ref{s:4wave}. In both cases the \nlcnls\ equations for the regime with $(\vg{1}-\vg{2})=O(1)$ lead to quadratic dispersion relations for the disturbance frequency, enabling simple stability  conclusions to be reached. One interesting feature of these results is that a coupled four-wave solution can be stable even whilst one or more of the four component waves can be unstable in isolation. This is in contrast to coupled two-wave solutions, for which stability is equivalent to that of the two component waves in isolation. For the regime with $(\vg{1}-\vg{2})=O(\eps)$, the \nlcnls\ equations lead to quartic dispersion relations for the disturbance frequency. The \cnls\ equations lead to a quartic dispersion relation for two-wave solutions, and to an eighth-order dispersion relation for four-wave solutions, although the latter was not evaluated here. 

Numerical methods were used to solve the fully nonlinear \sg\ equations (\ref{1.2}) in a periodic domain, using plane-wave initial conditions, as described in Section~\ref{s:numerics}. 
A major issue was to assess the performance of the stability predictions obtained from the different weakly nonlinear models.  
For cases with $(\vg{1}-\vg{2})=O(1)$, the quadratic dispersion relations of the \nlcnls\ equations give good predictions: relative errors are about 2\,\% when the amplitude parameter $\amp=0.5$, and these decrease with $\amp$. For two-wave cases, the quartic dispersion relation of the \cnls\ equations gives no more accurate predictions, even at large amplitudes. 
As $|\vg{1}-\vg{2}|$ decreases, the quadratic dispersion relations become less accurate, and the quartic dispersion relations of the \nlcnls\ regime with $(\vg{1}-\vg{2})=O(\eps)$, or equivalently of the \cnls\ equations in the two-wave case, become more useful. For cases where the group velocities are very close or equal, the quadratic dispersion relations of the \nlcnls\ equations no longer even predict qualitatively the correct form of the dispersion curve, and a quartic dispersion relation must be used. The region of maximum growth is then well described, with relative errors decreasing with $\amp$, although at larger disturbance wavenumbers convergence appears to be poor and relative errors are about 10\,\%. 
In some four-wave cases there is an additional dominant short-wavelength instability, which does not correspond to a long spatial modulation of the type considered theoretically, and is not predicted by any of our theories. 

The numerical simulations also enabled us to examine the long-time evolution of modulational instabilities. For the energy exchange parameters of interest, the initial solutions are spatially uniform plane-waves with an almost complete energy exchange between the components, whilst later in the evolution the solutions can become spatially localized with an incomplete energy exchange. Further, when the system is close to the integrable case at $c=1$, the time evolution is distinguished by a remarkable almost periodic sequence of energy exchange scenarios, with spatial patterns alternating between approximately uniform wavetrains and localized structures. As $c$ moves away from 1, the time evolution becomes more disordered. 

Although the whole of this study was performed for a two-component set of coupled \sg\ equations, we would expect similar phenomena to arise in a wide class of nonlinear wave systems. Thus, in addition to applications to multi-component systems such as those mentioned in the Introduction, processes of the type considered here should arise in other physical systems which can support several wave modes. For instance, an interesting example might be energy exchange between nonlinear internal waves in a layered, or continuously stratified, fluid.


\section*{Acknowledgments}

We are thankful to  E.S. Benilov, E. Knobloch and A.V. Mikhailov for useful and stimulating discussions.

\end{document}